\documentclass[fleqn,10pt]{wlscirep}
\usepackage{amsmath}
\usepackage{amssymb}
\usepackage{graphicx}
\usepackage{cite}
\usepackage{color}
\usepackage{etoolbox}  % patching other commands
\usepackage{outlines}
\usepackage{subcaption}
\usepackage{comment}
\usepackage{lineno}
\newcommand{\inhead}[1]{\textbf{#1}}

\newcommand{\fixmecolor}{blue}

\let\oldoutline\outline
\def\outline{\oldoutline\color{\fixmecolor}}

\newcommand{\teamnote}{\textcolor{green}}
\newcommand{\Jnote}{\textcolor{orange}}

\title{Time Series Methods and Ensemble Models to Nowcast Dengue at the State Level in Brazil}
\author[1]{Katherine Kempfert}
\author[2,3]{Kaitlyn Martinez}
\author[4]{Amir Siraj}
\author[5]{Jessica Conrad}
\author[2]{Geoffrey Fairchild}
\author[6]{Amanda Ziemann}
\author[2]{Nidhi Parikh}
\author[7]{David Osthus}
\author[8]{Nicholas Generous}
\author[2, *, +]{Sara Del Valle}
\author[2, +]{Carrie Manore}

\affil[1]{Department of Statistics, University of California Berkeley, Berkeley, CA, USA}
\affil[2]{Information Systems and Modeling group (A-1) of Analytics, Intelligence, and Technology Division, Los Alamos National Laboratory, Los Alamos, NM, USA}
\affil[3]{Department of Mathematics \& Statistics, Colorado School of Mines, Golden, CO, USA}
\affil[4]{Department of Biological Sciences, University of Notre Dame, Notre Dame, IN, USA}
\affil[5]{Theoretical Biology and Biophysics group (T-6) of Theoretical Division, Los Alamos National Laboratory, Los Alamos, NM, USA}
\affil[6]{Intelligence and Space Research Division, Los Alamos National Laboratory, Los Alamos, NM, USA}
\affil[7]{Statistical Sciences Group of Computer, Computational, and Statistical Sciences Division (CCS-6), Los Alamos National Laboratory, Los Alamos, NM, USA}
\affil[8]{Intelligence and Emerging Threats Program Office, Los Alamos National Laboratory, Los Alamos, NM, USA}

\affil[*]{corresponding author sdelvall@lanl.gov}
\affil[+]{these authors contributed equally}
\begin{document}

\begin{abstract}
%\instructions{Example Abstract. Abstract must be under 200 words and not include subheadings or citations.}

%In recent years, novel data streams such as Internet and social media data have been utilized in infectious disease models. has benefited from the inclusion of novel data streams such as Internet and social media data. 
Predicting an infectious disease can help reduce its impact by advising public health interventions and personal preventive measures. Novel data streams, such as Internet and social media data, have recently been reported to benefit infectious disease prediction. 
%Recent advances in data collection have allowed us to measure different factors contributing to disease dynamics such as emergent behaviors posted on social media or birds migration paths through satellite tracking. 
However, few efforts have quantitatively assessed the predictive benefit of novel data streams in comparison to more traditional data sources, especially at fine spatiotemporal resolutions. %However, sophisticated modeling systems have been underutilized for dengue, a ``neglected tropical disease". 
As a case study of dengue in Brazil, we have combined multiple traditional and non-traditional, heterogeneous data streams (satellite imagery, Internet, weather, and clinical surveillance data) across its 27 states on a weekly basis over seven years. For each state, we nowcast dengue based on several time series models, which vary in complexity and inclusion of exogenous data (seasonal autoregressive integrated moving average, vector autoregression, seasonal trend decomposition based on locally estimated scatterplot smoothing, and their variants). The top-performing model varies by state, motivating our consideration of ensemble approaches to automatically combine these models for better outcomes at the state level. %Our results vary by state and are correlated with several demographic and socioeconomic variables (e.g., education level). 
For a trimmed mean ensemble approach, 25 states achieve Pearson correlation coefficients (between fitted and observed values in the 2015-16 testing window) exceeding 80\%; meanwhile, the median value over the 27 states is 91.75\%, and the maximum is 96.44\%. Associated 95\% prediction intervals reach approximately 96\% empirical coverage or more for half the states. 
Model comparisons suggest that predictions often improve with the addition of exogenous data, although similar performance can be attained by including only one exogenous data stream (either weather data or the novel satellite data) rather than combining all of them. Among the Brazilian states, the model performance is spatially autocorrelated and associated with measures involving education, employment, and population. %Model comparisons suggest that exogenous data often contribute additional gains in nowcast accuracy, compared to when only past clinical surveillance data is used.
Our results demonstrate that Brazil can be nowcasted at the state level with high accuracy and confidence, inform the utility of each individual data stream, and reveal potential geographic contributors to predictive performance. Our work can be extended to other spatial levels of Brazil, vector-borne diseases, and countries, so that the spread of infectious disease can be more effectively curbed. 
%and demonstrate the value of integrating heterogeneous data streams. Our methodology and results can be extended to other spatial levels of Brazil, vector-borne diseases, and countries, so that the spread of infectious diseases can be more effectively curbed. 
%Our results can inform the timing and utility of each individual data stream and demonstrate the value of integrating heterogeneous data streams to enhance forecast accuracy. Our methodology and results can be extended to other spatial levels of Brazil, similar vector-borne diseases, and other countries. %Additionally, this works lays the groundwork for the extending forecasting methodologies to the municipality level.

\end{abstract}

\flushbottom
\maketitle

\thispagestyle{empty}

\section*{Introduction}

%\instructions{The Introduction section, of referenced text expands on the background of the work (some overlap with the Abstract is acceptable). The introduction should not include subheadings.}

In the past 15 years, dengue hemorrhagic fever and classic dengue have increased in many Latin American countries, including Brazil \cite{who_2015, cafferata2013dengue}. In addition to higher morbidity and mortality of dengue, the transmission pattern has changed from long intervals to annual outbreaks in multiple locations and persistent co-circulation of several serotypes \cite{who_2015, cafferata2013dengue, wilder2005serological, de2004dengue}. Understanding and accurately predicting dengue can improve mitigation efforts as well as inform modeling approaches for newly emergent mosquito-borne diseases. With over one million cases some years, Brazil as a country suffers among the highest numbers of dengue epidemics in the world \cite{guo2017global}. Despite its substantial global burden, dengue has received less attention than some other diseases (e.g., the flu) within the modeling community. We aim to address this gap by combining disparate data streams to nowcast dengue in Brazil at the state level. 

\begin{comment}
\begin{figure}[ht]{}
\includegraphics[width=0.75\textwidth]{Figures/Motivation_Timing.pdf}
\centering
\caption{\textbf{Dengue timing in Brazil, 2010-2016.} LEFT: Density plots showing the calendar week when states reach an early warning threshold (15\% of seasonal cumulative cases, TOP) and peak seasonal cases (BOTTOM) for each region as defined by the Brazilian government.  Each dot represents a state’s value from 2010--2016. The start of the season was determined separately for each region (North: September, Northeast: November, Central West: September, South: September, Southeast: October) based on the seasonal dynamics of the states in each region. The solid line represents the median timing of each region. RIGHT: Maps showing the month associated with the median calendar week of early warning (TOP) and peak thresholds (BOTTOM) across Brazil.}
\label{fig:timing}
\end{figure}
\end{comment}

Advances in computing, remote sensing, and modeling have enabled the creation and analysis of big data for public health applications\cite{dolley2018big, salathe2012digital, brownstein2009digital}. %In particular, big data have been exploited to understand and predict infectious diseases. Detecting current outbreaks (``nowcasting") and predicting future outbreaks (``forecasting") can help inform public health intervention and personal risk. Meanwhile, identifying key contributors to a disease can improve mitigation strategies. 
In particular, novel approaches have been developed that combine non-traditional, heterogeneous data streams with clinical surveillance data.
%to identify relationships and predict current or future disease spread. 
For example, several epidemiological studies have assessed the impact of climatic \cite{wu2007weather, lu2009time, johansson2009multiyear, colon2013effects, rosa2006associations, hii2012optimal, hii2012forecast, lowe2016evaluating, lowe2014dengue, shi2015three, buczak2012data, de2001use, viana2013ocurrence, campbell2015climate, gharbi2011time}, Internet, \cite{pollett2017internet, yang2017advances, chan2011using, milinovich2014using, althouse2011prediction, gluskin2014evaluation, priedhorsky2017measuring, althouse2015enhancing, gomide2011dengue, de2017dengue, guo2017developing}, and remote sensing data \cite{kalluri2007surveillance, mcfeeters2013using, louis2014modeling, nakhapakorn2005information, buczak2012data, machault2014mapping, laureano2017modelling, troyo2009urban, chang2009combining, hay1998predicting, kalluri2007surveillance, brown2008remotely, allen2006exploring, young2013remote, lacaux2007classification} for monitoring and forecasting disease.
To address challenges associated with novel data streams for disease surveillance, Althouse et al. have promoted a three-step system in a recent survey: ``(1) Quantitatively define the surveillance objective(s); (2) build the surveillance systems and model(s) by adding data (existing and novel) in until there is no additional improvement in model performance to achieve stated objectives, assessed by (3) performing rigorous validation and testing." \cite{althouse2015enhancing} For (2), few studies \cite{mcgough2017forecasting, santillana2015combining, majumder2016utilizing} have studied the relative contribution of each data stream when predicting disease; additionally, ``performance can be over-stated" when comparing models with novel data streams to ``trivial instances of traditional models" \cite{althouse2015enhancing}. Meanwhile, a number of studies \cite{gomide2011dengue, generous2014global, gluskin2014evaluation} have failed to satisfy (3), raising concerns about ``over-fitting" and inflated predictive performance. 

Our main contributions are as follows. 1) We combine traditional and nontraditional data sources (including remote sensing, Internet, climatological, and clinical surveillance data), which together have rarely been used for disease modeling and never for dengue modeling in Brazil. 2) For each of the 27 states in Brazil, weekly dengue is nowcasted over a six to seven year period; few disease modeling studies involving disparate data streams have attempted to model an entire country simultaneously at the state level. 3) For this task, we consider a selection of statistical models for time series: nontrivial baselines containing only past dengue case counts (seasonal autoregressive integrated moving average (SARIMA) and seasonal trend decomposition based on LOESS), variants of SARIMA that contain transformed and dimension-reduced variables from the diverse data streams, and a multivariate vector autoregressive model including transformed and dimension-reduced data. To automatically combine nowcasts, ensemble methods are also explored. 4) We compare cross-validated model performance for 2015-16 along several metrics (root mean square error (RMSE), relative RMSE, Pearson correlation, mean absolute error (MAE), and relative MAE); cross-validation mirrors practical nowcasting by training with data that would be available at that time. 5) We refit our models for subsets of data to quantitatively assess the benefit of each individual data stream versus their combination. Overall, we produce an accurate nowcasting system at the state level of Brazil and establish a spatiotemporal framework that does not assume one model for the entire nation. Beyond dengue in Brazil, we followed the standards outlined by Althouse et al. \cite{althouse2015enhancing} by empirically demonstrating how nowcasting is affected by model, space, and inclusion of certain data streams.

\subsection*{Dengue \& mosquito-borne disease modeling}

Recently, mathematical \cite{andraud2012dynamic}, statistical \cite{lowe2016evaluating, lowe2014dengue, lowe2013development, hii2012optimal, hii2012forecast, guo2017developing, de2004dengue, buczak2012data, gluskin2014evaluation,  shi2015three, yang2017advances}, climate-driven \cite{campbell2015climate}, and computational \cite{tegner2009computational, de2001use} models have been used to understand, simulate, and predict infectious diseases, particularly vector-borne diseases \cite{pollett2017internet, althouse2011prediction}. In our study, we consider models of dengue from statistics and machine learning, so we briefly summarize trends among similar work. Often, dengue has been modeled with traditional distributions for count data from statistics. For example, Quasi-Poisson \cite{hii2012optimal, hii2012forecast} and negative binomial distributions have been applied \cite{lowe2013development, guo2017developing, de2017dengue} because they account for over-dispersion in count data. In other cases, dengue has been modeled as a continuous quantitative variable \cite{gluskin2014evaluation} or a categorical risk level (e.g., high risk vs. low risk) \cite{buczak2012data}. Many studies have utilized methods not adapted to time series data (e.g., linear regression \cite{gluskin2014evaluation, generous2014global, chan2011using}), while others have explicitly accounted for temporal dependency through time series models (e.g., autoregression \cite{johansson2016evaluating, promprou2006forecasting, luz2008time, eastin2014intra, silawan2008temporal}). Traditional models from statistics have been explored (e.g., linear regression \cite{gluskin2014evaluation, generous2014global, chan2011using}, generalized linear models \cite{guo2017developing}, and SARIMA \cite{johansson2016evaluating, promprou2006forecasting, luz2008time, eastin2014intra, silawan2008temporal}), while methods from statistical and machine learning have also been deployed in recent years (e.g., tree-based methods \cite{guo2017developing}, support vector machines \cite{guo2017developing}, rule-based learning \cite{buczak2012data}, and least absolute shrinkage and selection operator (LASSO) regression \cite{shi2015three}). For prediction, various performance metrics have been reported, including correlation-based measures (e.g., coefficient of determination $R^2$ and correlation coefficient $R$) \cite{guo2017developing, yang2017advances, guo2017developing, hii2012optimal, gluskin2014evaluation, generous2014global, johansson2016evaluating, eastin2014intra} and error estimates (e.g., root mean square error (RMSE) and relative RMSE \cite{guo2017developing, yang2017advances, guo2017developing, luz2008time}, root mean squared percentage error \cite{yang2017advances}, mean absolute percentage error (MAPE) \cite{yang2017advances, shi2015three, silawan2008temporal}, and mean absolute error (MAE) and relative MAE \cite{yang2017advances, de2017dengue, johansson2016evaluating, eastin2014intra}).
%, and information criteria (e.g., Akaike's information criterion (AIC) and the Bayesian information criterion (BIC)). 
For classification, performance metrics include classification accuracy \cite{rahmawati2016using}, sensitivity \cite{buczak2012data, rahmawati2016using}, specificity \cite{buczak2012data, rahmawati2016using}, and area under the curve \cite{lowe2013development}. Lowe et al. have highlighted several issues in dengue prediction studies, including a lack of proper validation \cite{lowe2013development}; indeed, Althouse et al. have confirmed that only 41\% of selected papers on novel data streams contained any form of validation \cite{althouse2015enhancing}. 
Cross-validation or out-of-sample testing are essential for estimating true predictive performance and, hence, establishing the utility of a model.  
%Although our models use similar approaches as previous papers, we 

%and, hence, establishing the predictive performance of a model.%So and so et al. have emphasized a need for studies to include measures from epidemiology, such as x, as well. 

\subsubsection*{Internet data streams for disease applications}

A number of studies have analyzed the potential use of Internet data streams, such as Twitter, Google, and Wikipedia, to inform disease surveillance \cite{althouse2015enhancing, pollett2017internet, yang2017advances, chan2011using, milinovich2014using, althouse2011prediction,  gluskin2014evaluation, priedhorsky2017measuring}, since people use the Internet to search and share health-related information. For example, before visiting a doctor's office, people experiencing fevers, headaches, and other symptoms might search ``dengue symptoms" on Google, and tracing these searches could provide an early warning of an outbreak. %The premise is that people use the Internet to search and share health related information and that these digital traces can be captured through data extraction and analysis. 
Comprehensive surveys have been written \cite{althouse2015enhancing, pollett2017internet}, but here we highlight some studies focusing on dengue, particularly those in Brazil or using similar methods.
Althouse et al. analyzed Google search terms and dengue incidence data from Singapore and Bangkok and concluded that Google search terms helped predict dengue with high accuracy (out-of-sample $R^2$ up to $0.948$) \cite{althouse2011prediction}. Chan et al. used Google search query volume for dengue-related queries in Brazil, among other countries; for the national level of Brazil, they reported an extremely high correlation coefficient of $R=0.99$ in a holdout set \cite{chan2011using}.
%In later work, Gluskin et al. \cite{gluskin2014evaluation} predict dengue at the national level and state level of Mexico; they report correlation coefficients ranging from 0.01 to 0.88 for a selection of states, but there is no mention of validation. 
Gomide et al. identified a strong correlation of 0.9578 between tweets from Twitter and dengue surveillance data in Brazil, although there was no cross-validation or out-of-sample validation \cite{gomide2011dengue}. Generous et al. used Wikipedia access logs and clinical surveillance data to monitor and forecast dengue in Brazil and Thailand; while they reported an $R^2$ value of 0.85 for the nation of Brazil, their work included no cross-validation or out-of-sample validation \cite{generous2014global}. %Milinovich et al. \cite{milinovich2014using} used Internet search queries in Australia and compared them against clinical surveillance data for dengue; they demonstrated a significant correlation but did not assess predictive ability.
De Almeida et al. nowcasted and forecasted dengue in Brazil at the national and city level using tweets; at the national level, they reported adjusted $R^2$ ranging from 0.94 (for nowcasting) to 0.87 (for forecasting up to 8 weeks in advance) \cite{de2017dengue}.
%, while 33\% of cities suffered $R$ less than 0.5. 
Guo et al. have nowcasted and forecasted dengue within cities of a province of China based on Baidu search queries, as well as climate-related variables; support vector regression achieved a correlation coefficient of up to 0.999 in a 12 week forecast window \cite{guo2017developing}. Yang et al. predicted dengue at the national level of Brazil, among other countries; they reported $R=0.971$ in the recursive prediction window and demonstrated that inclusion of Google data improved model performance \cite{yang2017advances}. Finally, Infodengue continually combines Twitter data with climate and case count data to nowcast dengue in 788 cities across Brazil and reports the information online (\url{https://info.dengue.mat.br}) \cite{codeco2016infodengue, codeco2018infodengue}.

In addition to the lack of validation in several studies \cite{gomide2011dengue, generous2014global, gluskin2014evaluation} and focus on the national level in most studies \cite{althouse2011prediction, chan2011using, gomide2011dengue, generous2014global, milinovich2014using, yang2017advances}, few consider the contribution of Internet data to dengue prediction when past dengue count is already included in the model. In some countries, dengue clinical case data is limited, so surveillance efforts based only on Internet data are valuable. However, in Brazil, we assume good availability of clinical case data (up to a two week lag in reporting), so we believe the utility of Internet data should be judged against models already containing case data. With the exception of two studies \cite{de2017dengue, yang2017advances}, none of the selected studies on dengue in Brazil compare the performance of models containing Internet data to models already containing lagged case data. Similarly to Hii et al. \cite{hii2012forecast}, we develop models under the fundamental time series assumption that past behavior is the greatest influence of future behavior \cite{bowerman1993forecasting}. All our models contain past dengue cases and are quantitatively compared to models separately adding each exogenous data stream (out of Google Health Trends, satellite, and climatological/weather) then combining all data streams.

\subsubsection*{Satellite imagery for disease applications}

Traditional epidemiology can benefit from collection and analysis of satellite imagery \cite{hay2000overview, rogers2000remote}, since environmental factors influencing health can be identified remotely. Due to the difficulty of obtaining mosquito count data across time and space, satellite data have been used as proxies of mosquito habitat suitability and mosquito density \cite{kalluri2007surveillance, mcfeeters2013using}, especially in studies of dengue \cite{buczak2012data, machault2014mapping, nakhapakorn2005information, laureano2017modelling, troyo2009urban, chang2009combining}, malaria  \cite{hay1998predicting}, West Nile virus \cite{brown2008remotely, allen2006exploring, young2013remote}, and Rift Valley Fever \cite{lacaux2007classification}. For example, mosquitoes primarily subsist on vegetation, so the normalized difference vegetation index (NDVI), a satellite-derived measure of live green vegetation, can signal good mosquito habitats and high mosquito counts. We summarize several studies of dengue that are similar in scope to ours and note that dengue surveillance studies involving satellite data are limited (compared to studies exploiting weather and Internet data streams). Laureano et al. have measured sea surface temperature (SST), temperature, precipitation, and humidity variables, which account for 42\% of variation in dengue cases in a Mexican state \cite{laureano2017modelling}. For an area of Costa Rica, Troyo et al. have collected low, medium, and high resolution satellite imagery and explored potential relationships between NDVI, urban structural variables, and dengue cases \cite{troyo2009urban}. Buczak et al. have included measures of temperature, rainfall, altitude, NDVI, EVI, southern oscillation index, and sea surface temperature anomaly (SSTA), as well as demographic and political stability data to forecast dengue (as binary risk) for six districts in Peru; one classifier could detect outbreaks with 58.3\% accuracy and non-outbreaks with 96.1\% accuracy \cite{buczak2012data}.

A commonality of the selected studies is the small considered geographic scale: one state \cite{laureano2017modelling}, one area \cite{troyo2009urban}, and six districts in a province \cite{buczak2012data}. In contrast, we have collected every Landsat and Sentinel image over Brazil across seven years, derived weekly local-level indicators of vegetation, water, cloud cover, and burned areas, and then aggregated by over 5000 municipalities. Our data covers the approximate 3.3 million square miles of Brazil, the fifth largest country in the world by area. Hence, the satellite data alone are over 13 terabytes, and the processed indices include over 32 million entries over the seven-year period across over four satellites for Brazil. Additionally, with the exception of Buczak et al. \cite{buczak2012data}, the discussed studies only explore potential relationships among satellite data and dengue, rather than reporting validated predictive results. We compare cross-validated nowcasting results for various statistical models and empirically assess the contribution of each data stream to prediction.

\subsubsection*{Weather data streams for disease applications}

``Health and climate have been linked since antiquity," \cite{national2001under} and infectious disease modeling has investigated this connection in countless studies. For dengue modeling in particular, no exogenous variables have been explored as extensively as weather, due to the effects temperature, precipitation, and related variables have on mosquito habitat and life cycles \cite{yang2009assessing, watts1987effect, hopp2001global}. For example, three of the four mosquito life stages occur in stagnant water, so relatively high precipitation can often correspond to a large presence and count of mosquitoes. There are several studies of dengue prediction using weather variables \cite{wu2007weather, lu2009time, johansson2009multiyear, colon2013effects, rosa2006associations, hii2012optimal, hii2012forecast, lowe2016evaluating, lowe2014dengue, shi2015three, buczak2012data}, and the interested reader may refer to Viana et al. \cite{viana2013ocurrence} for a systematic review and to the World Health Organization for recent guidelines \cite{world2005using}. Here, we describe a selection of studies aligning most closely with our goals. Hii et al. included mean temperature and rainfall, as well as past dengue cases, in a model to predict dengue in Singapore; they reported training $R^2$ of 0.84, as well as good ability to classify outbreaks in the testing set \cite{hii2012forecast}. Lowe et al. forecasted dengue in Brazil based on temperature and precipitation data, plus population density, altitude, and past dengue relative risk, in a spatiotemporal hierarchical Bayesian model; three-month-ahead forecasts for June, 2014 achieved a hit rate of 57\% \cite{lowe2016evaluating, lowe2014dengue, lowe2013development}. As previously mentioned, Buczak et al. combined temperature and weather variables with demographic, political, and other environmental variables to forecast dengue in Peru using rule-based learning \cite{buczak2012data}. Shi et al. forecasted dengue in Singapore based on temperature and humidity, in addition to past dengue case counts and demographic information, and reported mean absolute percentage error (MAPE) ranging from 17\% (for one week forecasts) to 24\% (for three month forecasts) \cite{shi2015three}. As referenced previously, Guo et al. combined weather data (mean temperature, relative humidity and rainfall) with Internet data to nowcast and forecast dengue in the Chinese province Guangdong \cite{guo2017developing}. 

While traditional dengue surveillance and these selected studies indicate the importance of weather variables, few studies have quantified the combined contribution of weather data with novel data streams. In our study, we integrate the novel Google Health Trends and satellite data streams with the weather data (summary statistics of temperature, relative humidity, and daily temperature range) to nowcast dengue for all 27 states of Brazil. We quantify the contribution of each individual data stream by refitting our statistical models with subsets of the data and comparing the cross-validated performance across model and data subset. However, unlike Buczak et al. \cite{buczak2012data}, we do not use time series relating to political stability and demographics, since only 2010 yearly census data are available for Brazil.

\section*{Results}
%\instructions{Up to three levels of \textbf{subheading} are permitted. Subheadings should not be numbered.}

\subsection*{Model Comparison}
Literature review and an exploratory analysis (Figures \ref{fig:satellite}, \ref{fig:weather}, \ref{fig:GHT}, \ref{fig:cases_per_person}, and \ref{fig:cases_avg}) reveal spatial and temporal variation in each of the variables from diverse data streams (out of dengue clinical case count, Google Health Trends, satellite, and weather data). Such spatiotemporal heterogeneity through 2010-16 in Brazil has motivated consideration of several time series models, varying in complexity and inclusion of certain variables: seasonal autoregressive integrated moving average (SARIMA), SARIMAX with variables transformed through principal component analysis (PCA), SARIMAX with variables transformed through partial least squares (PLS), multiplicative and additive seasonal trend decomposition based on LOESS (STL), and vector autoregression (VAR) with variables transformed through PCA. We believe SARIMA, SARIMAX, STL, and VAR represent a reasonable selection of well-established time series models from statistics: SARIMA and STL contain only past dengue cases, while SARIMAX and VAR additionally contain the exogenous data streams, so we can fairly test the contribution of integrated Internet, satellite, and weather data for dengue nowcasting. For each state, each of these models is fit in the training weeks (2010-14) and recursively nowcasted for the testing weeks (2015-16). Our approach mirrors practical nowcasting, by accounting for the potential two week lag in dengue case reporting and utilizing an expanding training window. For real-time implementation, we need an automated method to generate good nowcasts for each state. To address this need, we explore ensemble approaches that combine the nowcasts: the first approach computes the trimmed mean, while the second approach calculates a weighted mean, where weights are chosen to be proportional to least absolute error in the training set.

\begin{figure}[ht]{}
\includegraphics[width=1\textwidth,trim={4cm 0cm 0cm 0cm},clip]{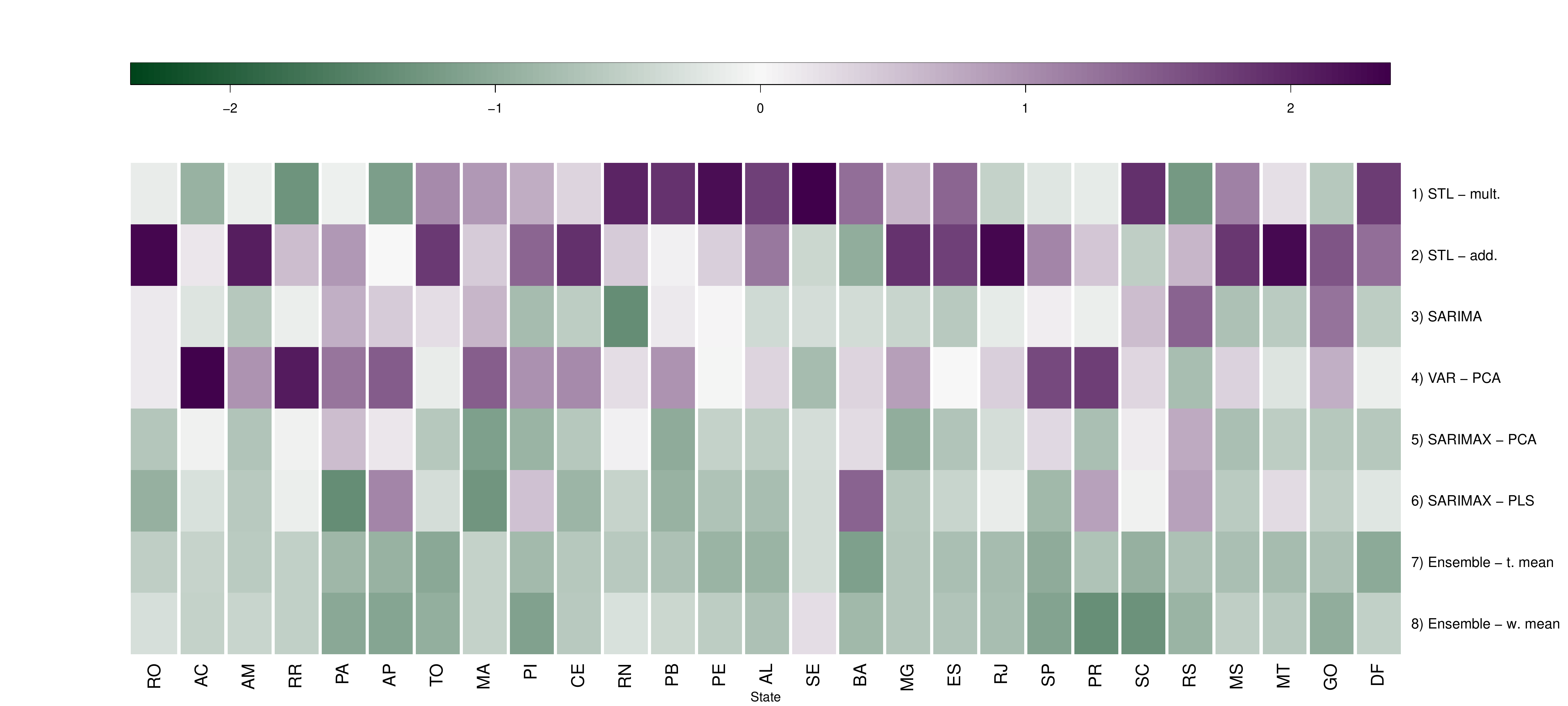}
\centering
\caption{\textbf{Heatmap of relative mean absolute error (RMAE) by model and state.} Among six statistical models of dengue and two ensembles to combine the models, we compare testing RMAE from recursive nowcasts in each Brazilian state. For a fixed state, the displayed RMAE is standardized among the methods: dark purple corresponds to higher standardized RMAE (worse methods), while dark green colors indicate lower standardized RMAE (better methods).}
\label{fig:heatmap}
\end{figure}

\begin{figure}[ht]{}
\includegraphics[width=1\textwidth,trim={0cm 0cm 0cm 1cm},clip]{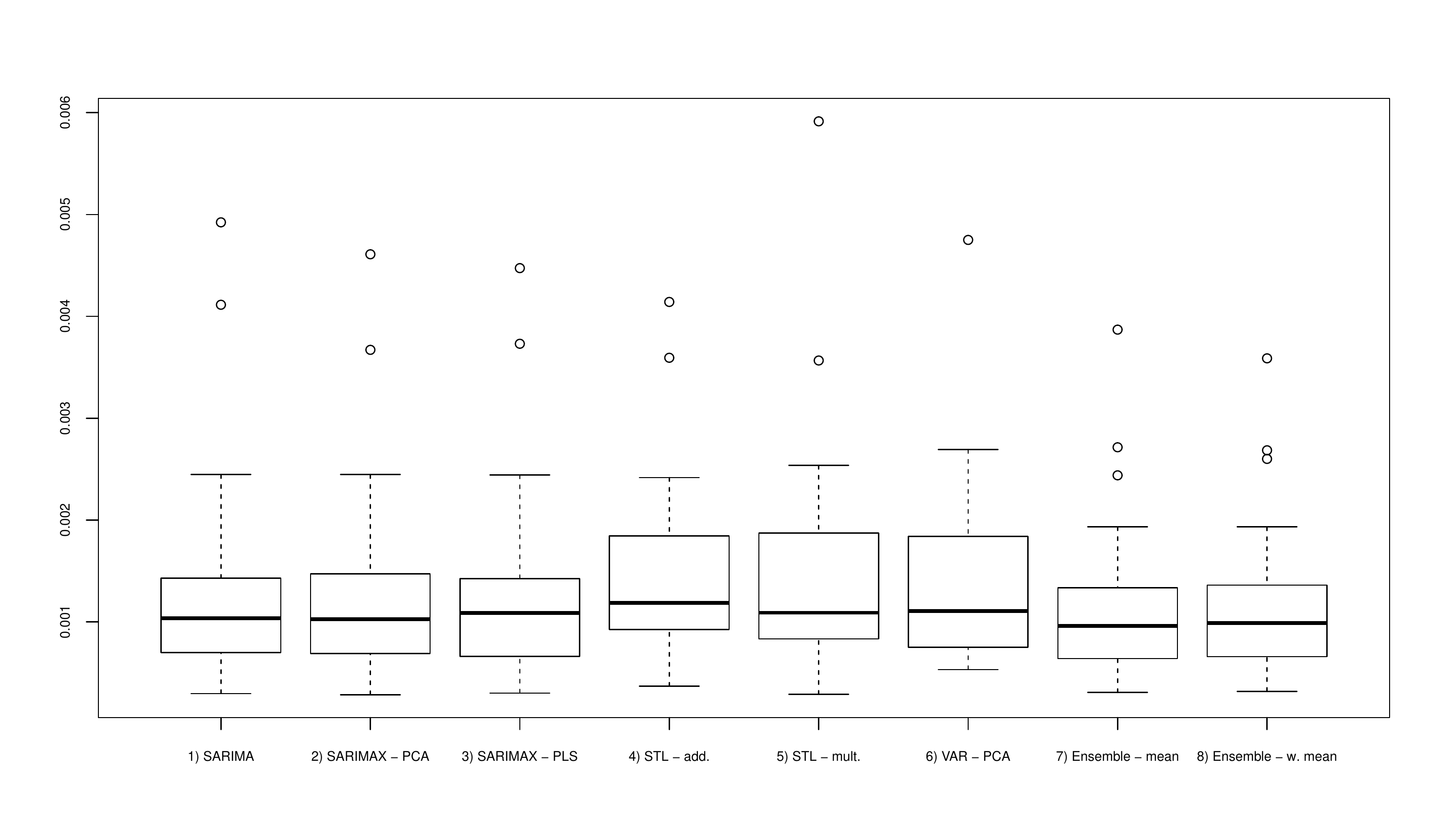}
\centering
\caption{\textbf{Side-by-side boxplots of relative mean absolute error (RMAE) by model.} Over the 27 Brazilian states, we compare the testing RMAE among six individual models of dengue and two ensemble methods that combine the individual models.}
\label{fig:boxplot}
\end{figure}

The relative mean absolute error (RMAE) of these methods (the six individual models plus two ensembles) is compared in Figures \ref{fig:heatmap} and \ref{fig:boxplot}, while numeric summaries for several performance metrics are presented in Table \ref{table:performance}. Each metric is computed within the testing set. Figure \ref{fig:heatmap} demonstrates that the ``best" model often varies by state, reinforcing the spatial heterogeneity of Brazil and the importance of separately modeling each state. Generally, the ensembles outperform the individual models, while the individual models SARIMA and SARIMAX (with PCA or PLS) outperform STL (additive or multiplicative) and VAR. For the RMAE performance metric, the ensembles yield optimum results in 15 states, while the individual models reach minimum error in 12 states. Of the individual models, SARIMAX (with PCA or PLS) attains minimum RMAE in seven states, while SARIMA, STL, and VAR are optimal for RMAE in only five states. Altogether, the models containing only past dengue count as a regressor reach minimal RMAE in only four states of the 27. These results indicate that the inclusion of exogenous data streams often improves dengue prediction, compared to when only past dengue counts are included in the model. The ensembles are shown to be helpful for improving average performance, e.g., the mean, median, and standard deviation of RMAE over the states are lowest for the ensembles. Similarly, the ensembles prevent bad nowcasts from occurring in a state, e.g., the maximum RMAE is lower for the ensembles than the individual models. In no cases have the ensembles produced the worst nowcasts; i.e., for each state and performance metric (out of MAE, RMAE, R, RMSE, and RRMSE), the highest error (or lowest R) are suffered by the individual models. For the trimmed mean ensemble method, Pearson correlations (between observed and predicted values in testing) are high, ranging from 0.733 to 0.964 with a median value of 0.918. We conclude that dengue can be accurately nowcasted for each state of Brazil, dengue prediction generally benefits from the inclusion of exogenous data streams, and nowcast performance can be improved by automatically combining models for each state through ensemble approaches.

\begin{table}[ht]{}
    %\centering
    \resizebox{\linewidth}{!}{
    \begin{tabular}{|l l|l|l|l|l|l|}
    \hline
        \textbf{Model} & &\textbf{MAE} &\textbf{RMAE} &\textbf{R} &\textbf{RMSE} &\textbf{RRMSE} \\
        \hline
          \textbf{SARIMA} &(Q0, Q2, Q4) &(6.20, 114.82, 2691.06) &(.002002, .00304, .008005) &(.678, .910, .956) &(8.70, 193.78, 5313.70) &(.00310, .00493, .0133)\\
                         &(Mean, SD) &(328.45, 601.70) &(.00348, .00127) &(.889, .0721) &(592.37, 1184.52) &(.00569, .00211)\\
         \textbf{SARIMAX - PCA} &(Q0, Q2, Q4) &(6.23, 105.47, 2784.62) &(.00198, .00294, .00714) &(.625, .920, .975) &(8.75, \textbf{177.79}, 6225.67) &(.00311, .00471, .0146)\\
                         &(Mean, SD) &(316.54, 591.57) &(.00334, .00116) &(.891, .0795) &(618.68, 1288.59) &(.00576, .00256)\\
         \textbf{SARIMAX - PLS} &(Q0, Q2, Q4) &(6.19, \textbf{104.16}, 2245.77) &(.00197, .00308, .00726) &(.618, \textbf{.923}, .970) &(8.69, 177.95, 4662.04) &(.000302, .00497, .0151)\\
                         &(Mean, SD) &(306.44, 525.27) &(.00336, .00116) &(.894, .0807) &(570.64, 1069.39) &(.00571, .00250)\\
         \textbf{STL - add.} &(Q0, Q2, Q4) &(6.65, 121.11, 3164.44) &(.00287, .00409, .00807) &(.563, .871, .936) &(9.61, 201.68, 6197.06) &(.00390, .00667, .0133)\\
                         &(Mean, SD) &(421.28, 787.53) &(.00419, .00134) &(.842, .0920) &(728.32, 1441.13) &(.00683, .00258)\\
         \textbf{STL - mult.} &(Q0, Q2, Q4) &(5.76, 126.07, 2532.55) &(.00241, .00370, .00717) &(.609, .891, .972) &(8.16, 205.97, 5221.25) &(.00369, .00616, .0108)\\
                         &(Mean, SD) &(363.97, 629.75) &(.00379, .00111) &(.876, .0807) &(683.30, 1302.39) &(.00647, .00205)\\
         \textbf{VAR - PCA} &(Q0, Q2, Q4) &(7.65, 121.61, 3429.73) &(.00223, .00362, .00631) &(.716, .881, .944) &(10.35, 196.05, 5906.13) &(.00334, .0553, .00954)\\
                         &(Mean, SD) &(394.66, 769.71) &(.00387, .00109) &(.878, .0598) &(640.03, 1304.89) &(.00602, .00180)\\
        \textbf{Ensemble - mean} &(Q0, Q2, Q4) &(5.94, 110.37, 2161.64) &(.00196, \textbf{.00275}, .00528) &(.733, .918, .964) &(8.36, 186.69, 4351.41) &(.00296, \textbf{.00459}, .00953)\\
                         &(Mean, SD) &(\textbf{291.03}, 510.72) &(\textbf{.00314}, .000901) &(\textbf{.906}, .0576) &(518.89, 995.13) &(\textbf{.00515}, .00159)\\
         \textbf{Ensemble - w. mean} &(Q0, Q2, Q4) &(5.94, 111.95, 2114.24) &(.00198, \textbf{.00275}, .00543) &(.697, .918, .969) &(8.40, 185.30, 4114.09) &(.00290, .00460, .00876)\\
                         &(Mean, SD) &(291.63, 504.01) &(.00318, .000905) &(.904, .0637) &(\textbf{512.22}, 955.75) &(.00518, .00160)\\

        \hline
    \end{tabular}
    }
    \caption{\textbf{Summary of nowcast performance over Brazilian states.} For each model paired with performance metric, summary statistics over the 27 states are presented for the testing weeks (2015-16). Q0, Q2, and Q4 correspond to the minimum, median, and maximum, respectively.}
    \label{table:performance}
\end{table}

\subsection*{Nowcast Results for Trimmed Mean Ensemble Approach}
\begin{figure}[htpb]{}
\includegraphics[width=1\textwidth,trim={0cm 0cm 0cm 0cm},clip]{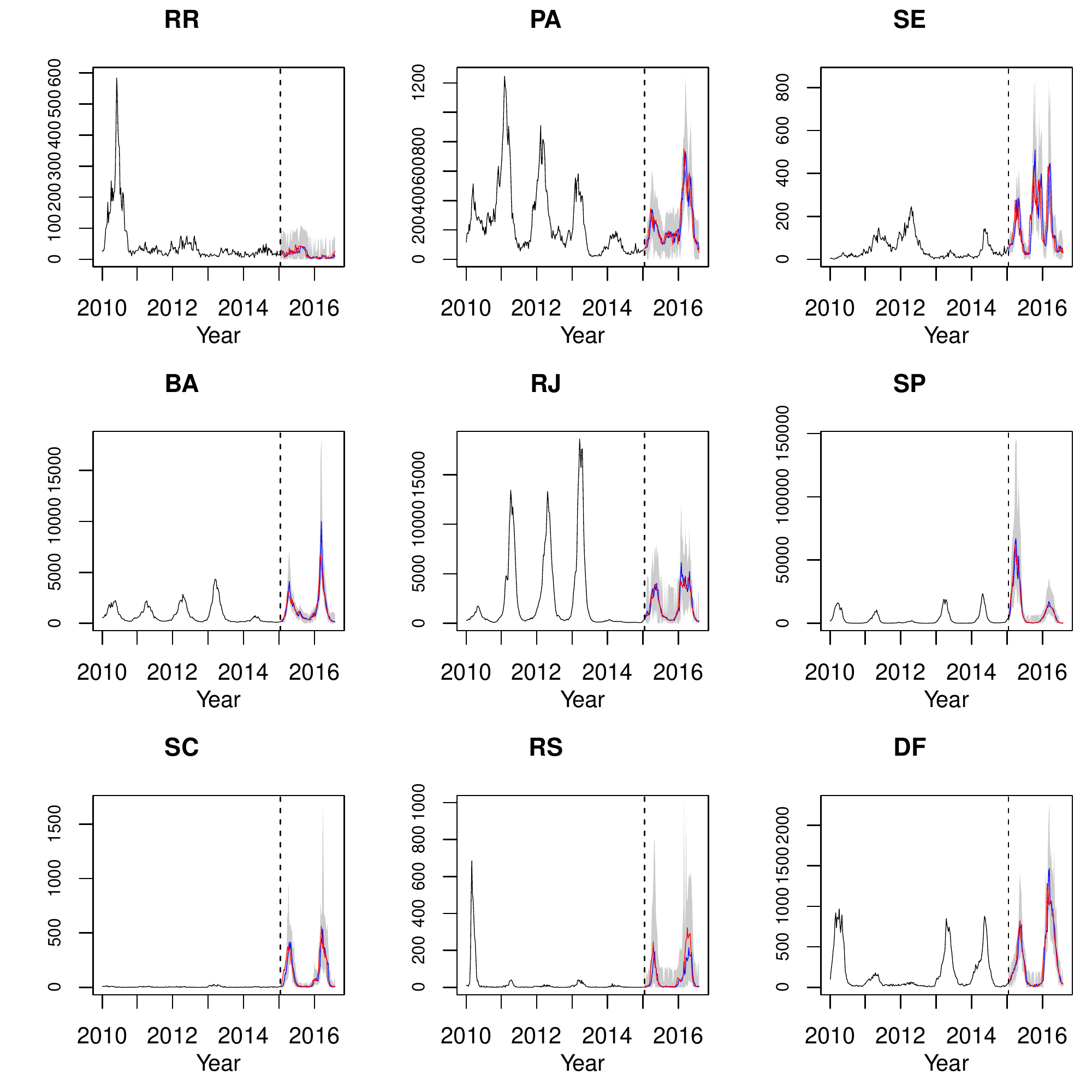}
\centering
\caption{\textbf{Observed vs. nowcasted dengue by Brazilian state from trimmed mean ensemble approach.} Time series of dengue are displayed for a selection of Brazilian states. %The model varies by state and is chosen from model selection within the training weeks. 
The vertical dashed line marks the boundary between training and testing weeks. The black time series corresponds to observed dengue in the training weeks (years 2010-14). The blue and red lines correspond to predicted and observed dengue, respectively, within the testing weeks (years 2015-16), while the gray shadings are the associated 95\% prediction intervals.} 
\label{fig:forecasts}
\end{figure}

Additional visualizations and performance metrics are provided for the trimmed mean ensemble approach, because of its relevance to practical nowcasting and generally good performance. %In the trimmed mean ensemble approach, nowcasts in the 0.1 and 0.9 quantiles are removed for each time step, then the mean is computed. Prediction intervals are generated through a conservative approach of
%the chosen statistical model varies by state and has been selected for each state as the one minimizing training RMSE. %In the trimmed mean ensemble approach, the chosen statistical model varies by state and has been selected for each state as the one minimizing training RMSE. 
As displayed in Figure \ref{fig:forecasts}, the observed and nowcasted dengue counts for the testing weeks (2015-16) are very close for the selected states (results for all states are included as supplementary materials). In many cases, the observed dengue count is captured within the 95\% prediction interval; empirical coverage for each state ranges from 81\% to 100\%, and the median and mean empirical coverage are approximately 96.3\% and 95.3\%, respectively. %Empirical coverage below nominal coverage is an expected consequence of applying inference procedures based on asymptotic normality. %and the approximation should improve in future work with greater case data availability (due to larger sample size). 
%%When a nowcast falls outside the prediction interval, it is more likely to be underpredicted than overpredicted here: over the states, the median proportion of underpredictions (i.e., the number of case counts above the prediction interval's upper bound divided by the total number of case counts falling outside the prediction interval) is approximately 0.725. 
Our strong results show that dengue within a state can be nowcasted with high accuracy and confidence.
%, although future work could address the prediction interval's empirical coverage and tendency to underestimate dengue counts.
%, and the approximation should improve with greater data availability (from increased sample size). When a dengue case count falls outside the prediction interval, it is more likely to be underpredicted than overpredicted These strong results indicate that dengue within a state can be nowcasted with high accuracy.

\begin{figure}[ht]
    \centering
    \begin{subfigure}[b]{.45\textwidth}
        \includegraphics[width=\linewidth]{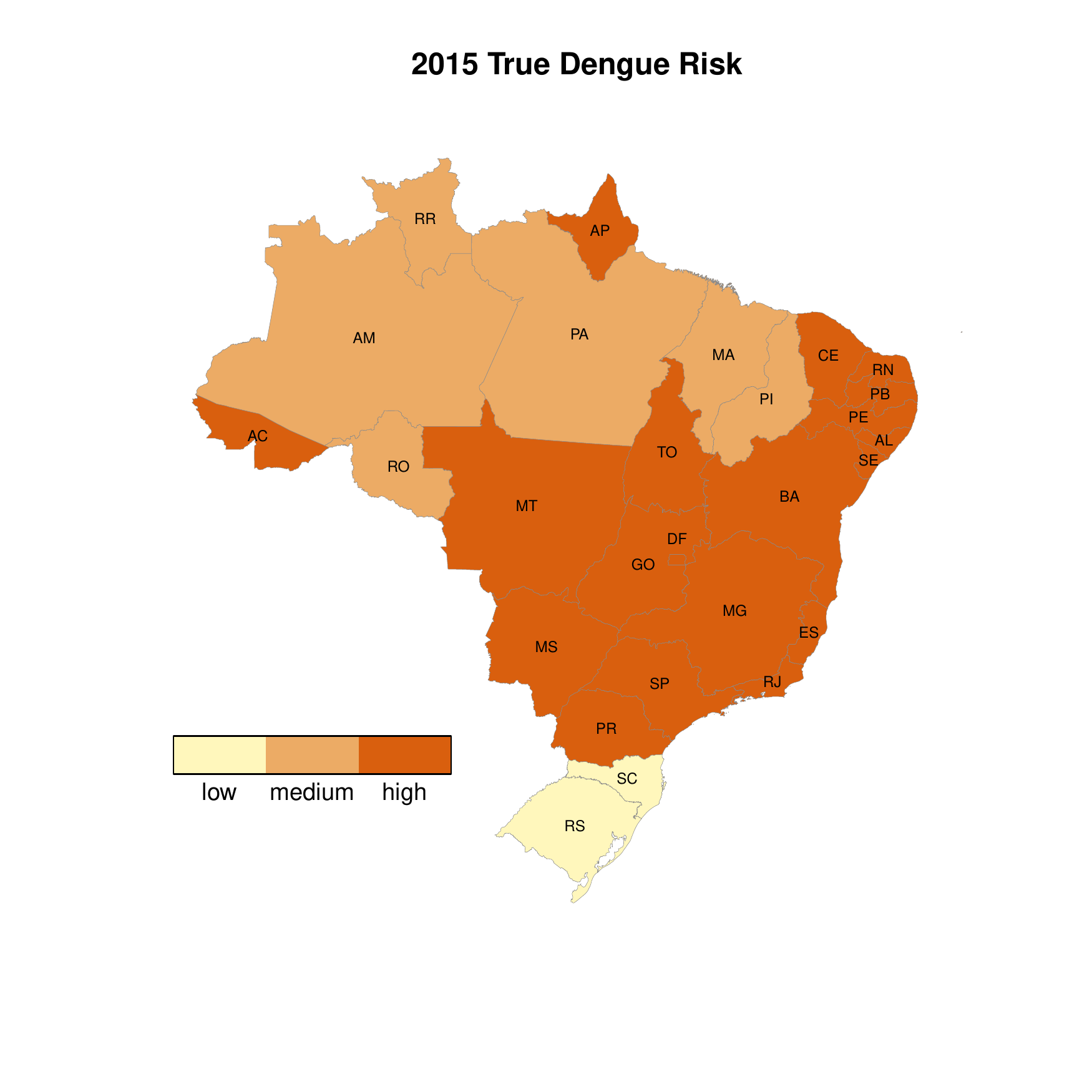}
        %\caption{}
        \label{subfig:true_risk}
    \end{subfigure}
    \begin{subfigure}[b]{.45\textwidth}
        \includegraphics[width=\linewidth]{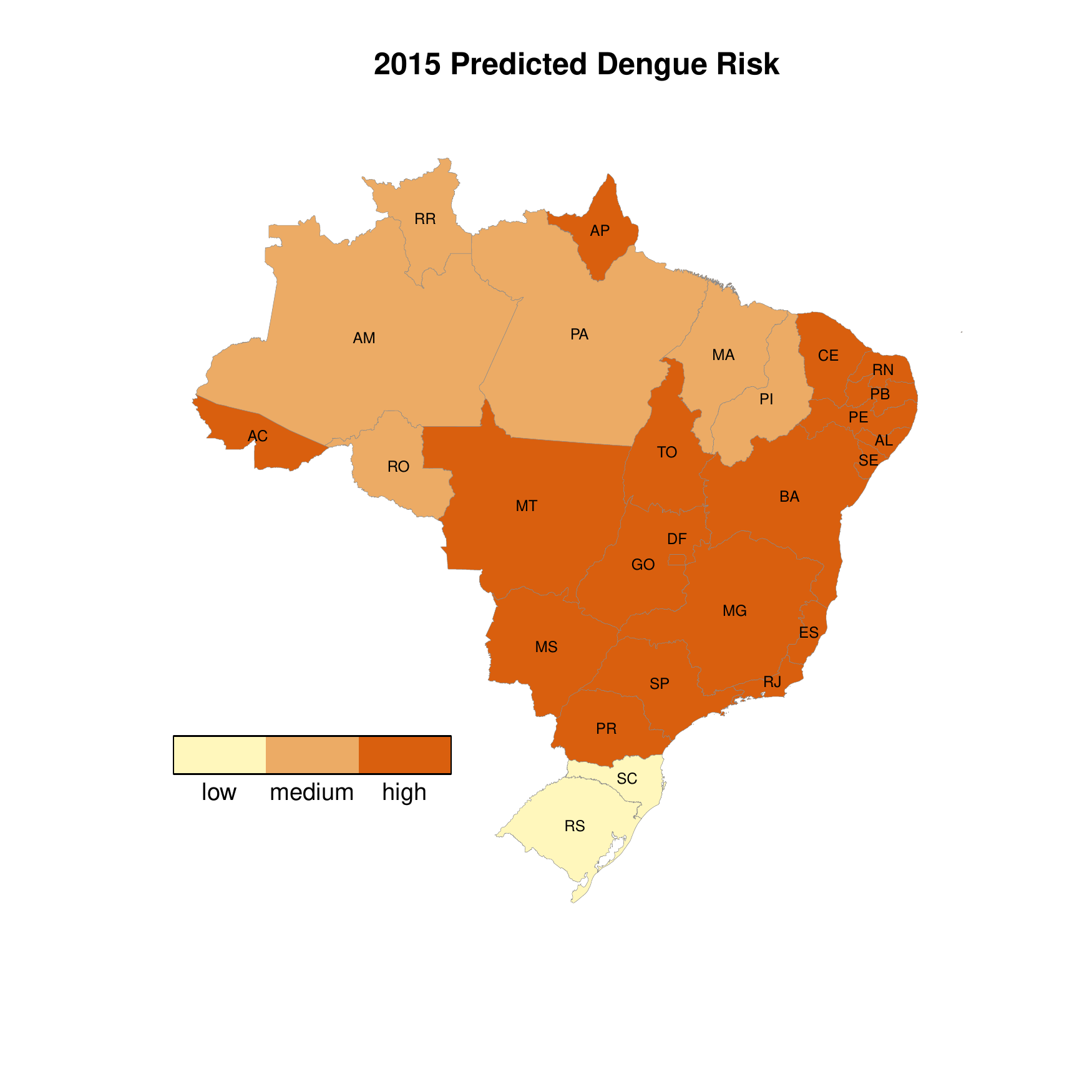}
        %\caption{}
        \label{subfig:pred_risk}
    \end{subfigure}
    \caption{\textbf{Observed vs. nowcasted dengue risk in Brazilian states for the trimmed mean ensemble approach.} Observed dengue risk and nowcasted dengue risk are displayed for the 2015 testing year in Brazil. In each state, low risk, medium risk, and high risk are respectively defined from the Brazilian Ministry of Health \cite{brazilian_ministry} as under 100, between 100 and 300, and more than 300 cases per 100,000 people per year. All 27 states are correctly classified.}
    \label{fig:risk}
\end{figure}

For improved application to public health decision-making, we assign the nowcasted dengue counts to risk categories. The National Dengue Control Programme of the Brazilian Ministry of Health defines low risk as less than 100, medium risk as between 100 and 300, and high risk as greater than 300 cases per 100,000 people in a year \cite{brazilian_ministry}. According to this definition, we convert the observed and nowcasted dengue case counts for each Brazilian state in 2015 to low, medium, or high risk (we do not consider 2016 here, since the dengue case counts are unavailable for the full year). Figure \ref{fig:risk} compares the observed versus predicted dengue risk and shows that all states' risk levels are correctly classified, further validating the performance of our approach.
%Such strong classification accuracy further validates the performance of our approach. 

\begin{figure}[ht]{}
\includegraphics[width=.8\textwidth, trim={0cm 3cm 0cm 2cm},clip]{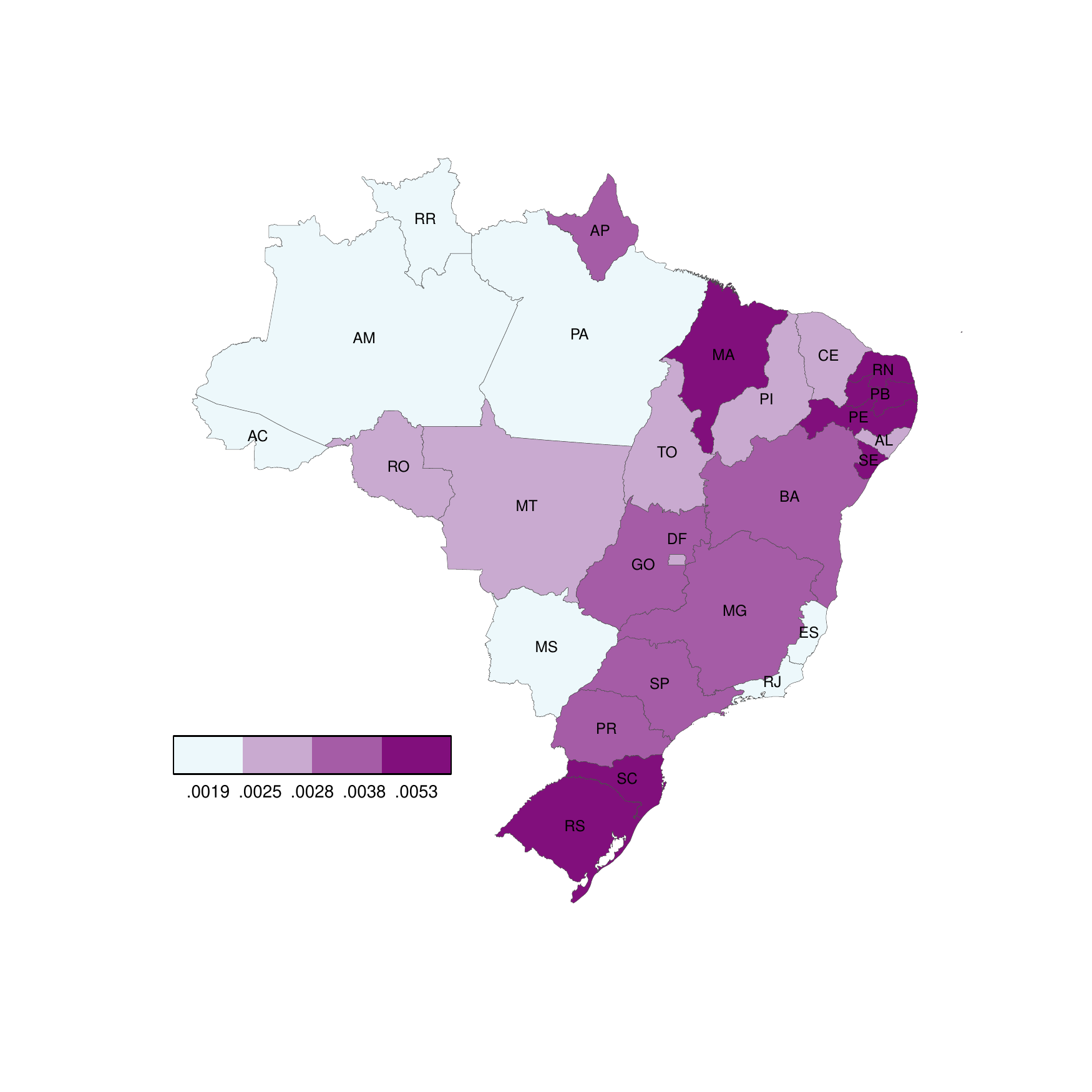}
\centering
\caption{\textbf{Relative mean absolute error (RMAE) over the Brazilian states from the trimmed mean ensemble approach.} Among the 27 Brazilian states, we compare the testing RMAE, where map breaks are chosen as quartiles. There is a low but statistically significant spatial autocorrelation in the states' error.}
\label{fig:error_map}
\end{figure}

\begin{figure}[htpb]{}
\includegraphics[width=1\textwidth]{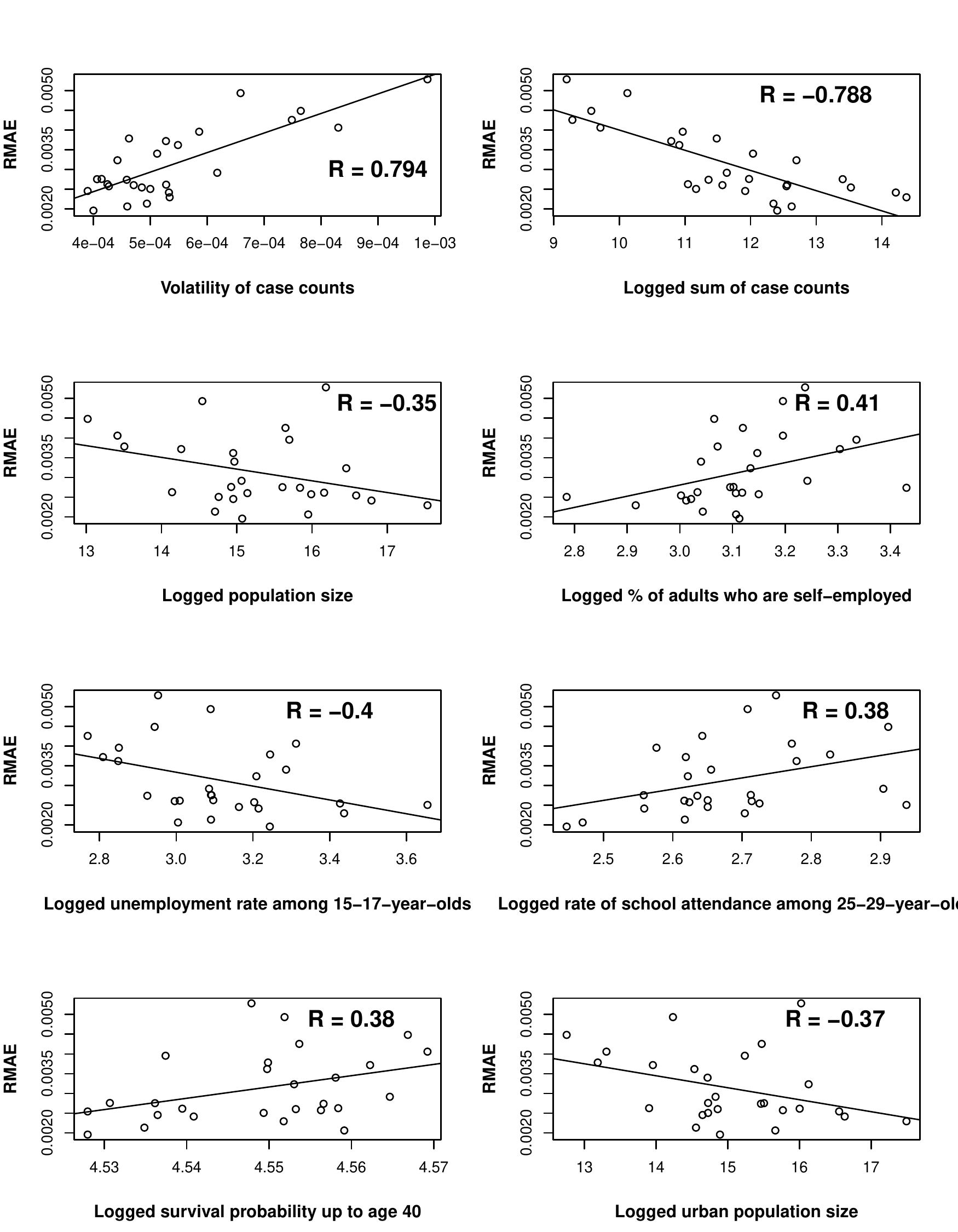}
\centering
\caption{\textbf{Correlating relative mean absolute error (RMAE) with variables across the Brazilian states.} We use census variables and features of the dengue case count time series (volatility and logged sum of case counts) to predict state RMAE through simple linear regression models. For each explanatory variable, the corresponding fitted line, observed points, and Pearson correlation coefficient $R$ are displayed.}
\label{fig:scatterplots}
\end{figure}

Figure \ref{fig:error_map} demonstrates that RMAE varies across the states; e.g., the states RS, SC, MA, RN, PB, PE, and SE have suffered the highest error, while the states AC, AM, RR, PA, and MS have achieved the lowest error. While the spatial autocorrelation of RMAE among adjacent states is relatively low (Moran's \cite{moran1950notes} $I = 0.21$), it is statistically significant with a $p$-value of 0.03 (from a Monte Carlo test assuming randomly distributed RMAE across Brazilian states). Regression analyses in Figure \ref{fig:scatterplots} reveal relationships among the states' RMAE, features of the dengue case count time series, and census variables relating to education, employment, health, and population. For each potential explanatory variable, we consider both its unlogged and natural logged form and report the strongest correlation of the two. For qualities of time series that could impact prediction, we examine sum of observations and define a measure of volatility. The sum and the volatility are respectively expressed as
\begin{equation}
\label{eq:case_vol}
     \sum_{t=1}^n y_t \textrm{  and  } \frac{\sum_{t=1}^n |y_t - y_{t-1}|}{n \sum_{t=1}^n y_t},
\end{equation}
where $y_t$ is the observed dengue case count for a state and $n=342$ is the number of observations from 2010 to 2016.
RMAE is most strongly correlated with the volatility ($R=0.794$), followed by the logged sum of case counts ($R=-0.788$). State error tends to increase as volatility of the case counts increases, which could be due to noise obscuring the underlying signal. Meanwhile, high error generally corresponds with low sums of case counts, suggesting that some states might suffer from small sample size issues. However, population size is not a strong predictor of volatility, sum of case counts, or RMAE ($R=-0.35$). The census variables explaining the most variation in state RMAE are logged percentage of adults who are self-employed, logged unemployment rate among the population aged 15 to 17 years, logged rate of school attendance among the population aged 25-29 years, logged survival probability up to age 40, and logged urban population size. Many census variables have nonobvious regression effects on RMAE. For example, we would generally expect a high unemployment rate in a state to contribute to higher error, due to potentially worse health infrastructure and access. In contrast, a high unemployment rate among minors might have the opposite effect, since unemployed minors might enjoy greater educational and wealth privileges that positively correlate with health infrastructure and access. Indeed, this latter interpretation is suggested by the negative correlation between RMAE and unemployment rate among 15-17-year-olds. Evidently, employment, education, health, urban population size, and qualities of the dengue case time series itself might affect nowcast performance in complex ways. %We have explored more sophisticated regression models of RMAE, including multiple linear regression and regularized regression, but

\begin{comment}
\begin{figure}[H]
\includegraphics[width=1\textwidth,trim={0cm 3cm 0cm 0cm},clip]{Figures/Categorical_Risk_Maps/Risk_Maps1.pdf}
\centering
\caption{Observed dengue risk and SARIMA-predicted dengue risk are displayed (by row) for the January and June months of 2015 (by column). In each state, low risk, medium risk, and high risk are respectively defined from the Brazilian Ministry of Health \cite{brazilian_ministry} as under 100, between 100 and 300, and more than 300 cases per 100,000 people. Beneath each month's column, the corresponding confusion matrix is presented, where non-high risk represents the low and medium risk categories.} 
\label{fig:risk_map1}
\end{figure}
\end{comment}

\subsection*{Assessing Utility of Individual Data Streams}
We empirically assess the benefit of each individual data stream to dengue nowcasting in Brazil, by refitting several statistical models with three subsets of data: 1) past case counts and Google Health Trends variables, 2) past case counts and satellite variables, and 3) past case counts and weather variables. The resulting error from each subset and model can help quantify the utility of each exogenous data stream (out of Google, weather, and satellite), when past dengue case counts are already included in the model. We consider SARIMAX with PCA, SARIMAX with PLS, and VAR with PCA (i.e., all the models able to incorporate such data) and nowcast dengue following the same protocol as above, where the only distinction is which data streams are included. 

\begin{figure}[htpb]{}
\includegraphics[width=1\textwidth]{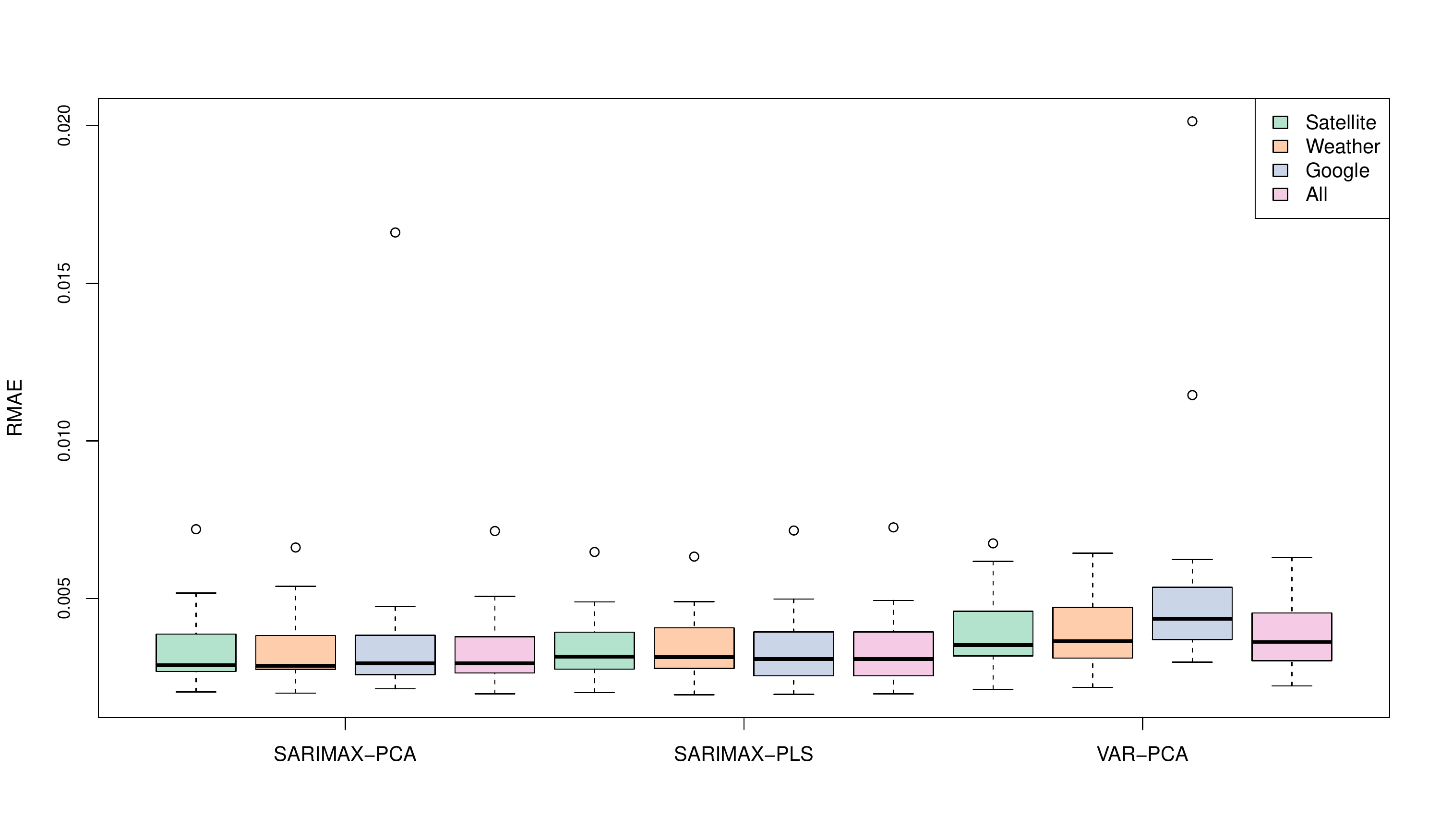}
\centering
\caption{\textbf{Quantifying the contribution of individual data streams vs. combination of data streams.} Relative mean absolute error (RMAE) across the states is compared for models containing subsets of data streams, e.g., the boxplots marked with ``Google" correspond to models fit using the Google Health trends data and past case counts, while boxplots marked with ``All" correspond to models fit using all exogenous data streams (Google, weather, and satellite) and past case counts.} 
\label{fig:datastreams_boxplot}
\end{figure}

Figure \ref{fig:datastreams_boxplot} displays the RMAE for each considered model paired with individual data stream, as well as previous results from combining data streams. %Additionally, we include RRMSE results from combining the three data streams (from the first results section). 
For the SARIMAX models (with PCA or PLS), the distribution of RMAE is very close for all data streams. For the VAR model, the satellite and weather data streams yield performance similar to the combination of datastreams, while the Google data suffer visibly higher RMAE and two severe outliers.

Next, we compare the models using individual data streams (SARIMAX and VAR) to models using no exogenous data (STL and SARIMA) and models combining the exogenous data (SARIMAX, VAR, and the ensembles). For 13 out of 27 states, minimum RMAE is attained when combining the exogenous data streams; for 9 of these 13 states, the ``best" model is an ensemble. Meanwhile, 10 states reach minimum RMAE for models using an individual data stream. Models involving no exogenous data achieve optimal RMAE in only four states (AC, RR, AP, and RS). Interestingly, multiplicative STL is the ``best" model for all four of these states, and three (AC, RR, and AP) of the four states are in the North region and represent the smallest state populations in Brazil. These results confirm that dengue nowcasting generally improves with the addition of exogenous data, although the utility of three exogenous data streams versus one differs by Brazilian state.

\section*{Discussion}
%\instructions{The Discussion should be succinct and must not contain subheadings. Explanation of how the results are awesome.}

We have integrated various traditional and novel data streams, including clinical case, weather, Google Health Trends, and satellite data to nowcast all 27 states of Brazil on a weekly basis from 2010 to 2016. Until the current work, these data streams have never been combined to model dengue in Brazil and have been rarely combined for other disease surveillance studies. %The Google Health trends data include 19 dengue-related keywords in English and Portuguese across the states of Brazil. The weather data, obtained from 613 weather stations, contain measures of temperature, relative humidity, and daily temperature range that are known to affect dengue transmission. The satellite data estimate each municipality's vegetation, water, cloud cover, and burned areas that can signal mosquito habitat suitability. The clinical case data are newly diagnosed cases per epidemiological week and municipality, which are considered ``gold standard" for mosquito-borne disease reporting. 
The data are big (e.g., 13 terabytes to store the satellite data) and heterogeneous in structure, necessitating computational resources and strategies such as high-performance, cloud, and parallel computing.
%to store, process, and aggregate thee d

Based on these high-dimensional, complex-structured, and spatiotemporally-distributed data, we have nowcasted dengue through methods from statistics and machine learning. For each state, we have fit six individual time series models: several good baselines containing only past dengue cases (SARIMA, multiplicative STL, and additive STL), two extensions of SARIMA using past dengue cases and transformed, dimension-reduced exogenous data (SARIMAX with PCA and SARIMAX with PLS), and a multivariate model of past dengue cases and transformed, dimension-reduced exogenous data (VAR with PCA). The model attaining minimum RMAE varies across the states, which has supported our decision to model at the state level and motivated methods to automatically generate good nowcasts for each state. For the latter, we have explored ensemble approaches from statistical and machine learning, including a trimmed and weighted mean of the individual model nowcasts. The median Pearson correlation coefficient (between observed and nowcasted values) over the states for the trimmed mean ensemble is 0.918. Meanwhile, the corresponding 95\% prediction intervals reached approximately a mean empirical coverage of 95\% over the states. The low error of point estimates, the excellent coverage of prediction intervals, and the application to practical, real-time nowcasting render our ensemble approach particularly promising. Performance comparison of the baselines versus other models confirm that the exogenous data sources (out of Google, weather, and satellite) offer additional information beyond past dengue case count. However, we have found that similar nowcast performance can be attained by using only one of the data sources (especially satellite or weather) in a model, rather than combining all three. This finding reflects the general yet counterintuitive fact that more data does not guarantee more accuracy. Regardless, our integration of these big, heterogenous data has generated very accurate nowcasts, which reach up to a 97\% Pearson correlation with observed dengue count in testing and enable 100\% classification accuracy of state risk level in 2015. 
%Finally, we have individually assessed the contribution of each exogenous datastream by refitting the SARIMAX and VAR models for three subsets of the data: past cases and Google, past cases and weather, and past cases and satellite. On average, the models combining the data streams have outperformed these refitted models, empirically demonstrating the value of integrating such data. Meanwhile, the models containing satellite variables generally achieved the best performance, followed by weather, then Google Health trends. 

%Such potential issues reflect the general yet counterintuitive fact that more data does not guarantee more accuracy. 

Through definition of several well-established performance metrics (RMSE, RRMSE, Pearson correlation coefficient $R$, MAE, and RMAE), empirical comparison of nontrivial models containing various subsets of data streams, and rigorous cross-validation that simulates practical nowcasting of dengue in Brazil, we believe we have met the standards outlined by Althouse et al. for integrating novel data streams into disease surveillance systems. We have established a flexible spatiotemporal framework for predicting dengue in Brazil that does not impose the assumption of a uniform model across space. Few studies have combined four or more data sources to predict disease across an entire nation at the state level or finer, and our work demonstrates that such modeling efforts are capable of obtaining strong predictive results. Further improvements are possible, particularly through dimension reduction techniques tailored to time series data, direct multi-step forecasting, and the addition of more diverse models in the ensemble. %(e.g., neural networks, $k$-nearest neighbor, and other nonlinear models from machine learning). 
Our next steps will involve extending our nowcasting system to forecasting at increased lead times and performing systematic comparison by model, lead time, and inclusion of certain data streams. Additionally, our methodology will be applied to the remaining spatial levels of Brazil, e.g., mesoregion, microregion, and municipality, to investigate how performance is affected by spatial granularity. Finally, methodology and results can be generalized to similar countries and mosquito-borne viruses. 

\begin{comment}
Intelligently exploiting such large, diverse data streams involves statistical and computational complexity, which is further amplified by modeling at finer spatial levels. The combination of these data streams is often large, high-dimensional, complex-structured, and spatiotemporally-distributed. Such complicated data require methods that can extract meaningful patterns, accurately predict disease, and provide validation through probabilistic estimates of error and variation. This motivates our consideration of the following statistical and machine learning approaches: time series models to describe and forecast dengue; dimension reduction and variable selection to identify succinct, meaningful subsets of covariates; cross-validation to accurately assess predictive performance; %inference procedures to assess significance of relationships between covariates and dengue at specific lags;
and systematic comparison of cross-validated model performance utilizing different data sources. 
%; and correlation between optimum model performance and demographic variables for the Brazilian states. 
The computational intensity of such methods, in conjunction with the size of the data, necessitate computational strategies and resources, including high-performance computing (HPC), cloud computing, and parallelism. 
\end{comment}

\section*{Methods}
%\instructions{Topical subheadings are allowed. Authors must ensure that their Methods section includes adequate experimental and characterization data necessary for others in the field to reproduce their work.}

In the Data section, we describe our data's source, its aggregation, and its exploratory analysis. The data exhibit strong seasonality, autocorrelation, spatial heterogeneity, and high-dimensionality, motivating our predictive modeling approach. In the Predictive Modeling Approach section, we discuss cross-validation, dimension reduction, and our spatiotemporal modeling framework. In the Individual Predictive Models section, we explain the considered time series models of dengue in Brazil. Finally, the Combining Predictive Models section details ensemble methods to automatically generate good nowcasts for each state. 

\subsection*{Data}
Over the seven year period from January 3, 2010 to July 17, 2016, we use as data streams multispectral satellite imagery, climatological data, and Google Health Trends to describe and predict clinical surveillance data of dengue count in Brazil. All data streams have been aggregated to the state level and weekly time unit for modeling, which results in up to 115 time series variables over 342 weeks for each state. When aggregating, we consider several summary statistics (minimum, mean, maximum, and standard deviation), as a compromise between retaining valuable statistical information and limiting the data's dimension.

\begin{table}[ht]{}
    \centering
    \begin{tabular}{|l|l|l|}
    \hline
        \textbf{Category} &\textbf{Source} &\textbf{Variables} \\
        \hline
          \textbf{Satellite} &\textbf{Descartes Labs} &\textbf{Normalized Difference Vegetation Index (NDVI)}\\
        & &\hspace{.5cm} Minimum, mean, and maximum (over 7 days) \\
        & &\hspace{1cm} Minimum, mean, maximum, and standard deviation (over municipalities)\\
       
        & &\textbf{Green Normalized Difference Water Index (NDWI)}\\
        & &\hspace{.5cm} Minimum, mean, and maximum (over 7 days) \\
        & &\hspace{1cm} Minimum, mean, maximum, and standard deviation (over municipalities)\\
       
        & &\textbf{SWIR Normalized Difference Water Index (NDWI)}\\
        & &\hspace{.5cm} Minimum, mean, and maximum (over 7 days) \\
        & &\hspace{1cm} Minimum, mean, maximum, and standard deviation (over municipalities)\\
       
          & &\textbf{SWIR Normalized Burn Ratio (NBR)}\\
          & &\hspace{.5cm} Minimum, mean, and maximum (over 7 days) \\
        & &\hspace{1cm} Minimum, mean, maximum, and standard deviation (over municipalities)\\
        
          & &\textbf{SWIR Proportion of Cloudy Pixels}\\
          & &\hspace{.5cm} Minimum, mean, and maximum (over 7 days) \\
        & &\hspace{1cm} Minimum, mean, maximum, and standard deviation (over municipalities)\\
      
        \hline
        \textbf{Weather} &\textbf{NOAA GSOD\cite{NOAA}} &\textbf{Temperature}\\
        & &\hspace{.5cm} Minimum, mean, and maximum (over 7 days) \\
        & &\hspace{1cm} Minimum, mean, maximum, and standard deviation (over municipalities)\\
        
        & &\textbf{Relative humidity}\\
        & &\hspace{.5cm} Minimum, mean, and maximum (over 7 days) \\
        & &\hspace{1cm} Minimum, mean, maximum, and standard deviation (over municipalities)\\
       
        & &\textbf{Daily Range in Temperature}\\
        & &\hspace{.5cm} Minimum, mean, and maximum (over 7 days) \\
        & &\hspace{1cm} Minimum, mean, maximum, and standard deviation (over municipalities)\\

        \hline
        \textbf{Google} &\textbf{Google Trends}\cite{google} &\textbf{``aedes"}\\
        & &\hspace{.5cm} ``aedes aegypti", ``aedes eg\'{i}pcio", ``aegypti", ``eg\'{i}pcio"\\
        & &\textbf{``dengue"}\\
        & &\hspace{.5cm}``dengue virus", ``dengue \'{e} v\'{i}rus", ``dengue fever",\\
        & &\hspace{.5cm}``dengue hemorrhagic fever", ``dengue sintomas", ``DENV", ``DHF"\\
        & &\textbf{``mosquito"}\\
        & &\hspace{.5cm} ``mosquito dengue", ``mosquitoes", ``novo v\'{i}rus da dengue", \\
        & &\hspace{.5cm} sintomas da dengue, v\'{i}rus da dengue\\
        \hline
        
        \textbf{Clinical Dengue Cases} &\textbf{Brazilian Ministry} &\textbf{Count of confirmed dengue cases}\\
        & \textbf{of Health}\cite{brazilian_ministry} &\\
        \hline
    \end{tabular}
    \caption{\textbf{Summary of data streams for modeling dengue in each Brazilian state.}}
    \label{table:data}
\end{table}

\subsubsection*{Satellite Imagery}
%\Jnote{Reference DesCartes Lab.}
For each municipality from 2010 to 2016, we have collected data via the Descartes Labs platform from four satellites (Sentinel-2 and Landsats 5, 7, 8) that imaged the full municipality each day. Various indices related to mosquito habitat (e.g., normalized burn ratio) are computed at the 10-30 meter resolution. To aggregate the daily values to weekly values, we compute over the seven days of each week several summary statistics: minimum, mean, and maximum. Next, these weekly variables, collected from the municipality level, are aggregated to the state level. For each state, summary statistics (minimum, mean, maximum, and standard deviation) of each weekly satellite variable are computed over the municipalities. This results in a total of 52 satellite variables per state, summarized in Table \ref{table:data}. For example, for the state of Minas Gerais (MG), the minimum, mean, maximum, and standard deviation of the mean NDVI variable are calculated over its 853 municipalities.

\begin{figure}[ht]
\includegraphics[width=1\textwidth]{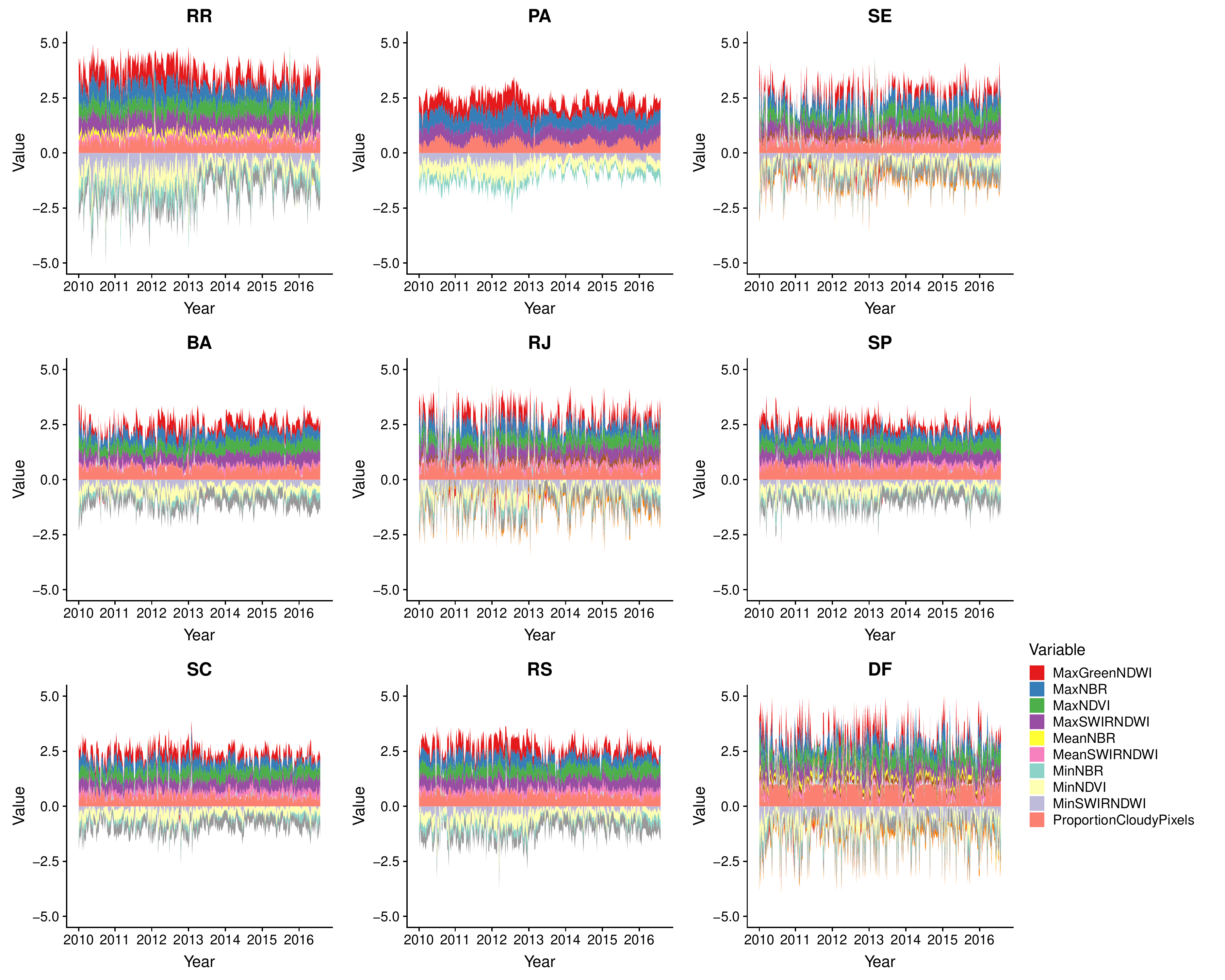}
\centering
\caption{\textbf{Satellite variables in each Brazilian state.} For a selection of Brazilian states, satellite variables are plotted from 2010 to 2016. (Variables with approximately zero variance in all states are omitted.) Plots for the remaining 18 states are included as supplementary materials.} 
\label{fig:satellite}
\end{figure}

For each state, the variables resulting from the mean summary statistic (over municipality) are plotted in Figure \ref{fig:satellite}. The satellite variables tend to reach more extreme values during the years 2010 to 2013, as compared to 2014-6; e.g., in Roraima (RR), the variables Minimum Green NDWI, Minimum NBR, Minimum NDVI, Minimum SWIR, and Mean Green NDWI reach lower values in 2010-13 (than in 2014-16), while the remaining variables achieve higher values in 2010-13 (than in 2014-16). Meanwhile, differences can be observed in the satellite variables between states. For example, the Mean NBR variable has exhibited approximately no changes over time (zero variance) in all shown states, except for RR, RJ, SC, and DF. Additionally, the Minimum SWIR NDWI variable reaches lower values in RR and PA than the other shown states. Such spatiotemporal distinctions between states' satellite variables may correspond to spatiotemporal differences in mosquito habitat, motivating our decision to model each state's dengue count with these satellite time series variables. 

\subsubsection*{Climatological Data}

We have collected data from weather stations via the “Global Surface Summary of the Day” (GSOD) dataset from the National Oceanic and Atmospheric Administration (NOAA) \cite{NOAA}. There are 613 weather stations across Brazil that provide daily climatological data. We select a subset of measures associated with dengue, including summary statistics of temperature and humidity. Since there are 5564 municipalities in Brazil (as of 2010), many localities are not covered by these stations. To address this limitation, we have used kriging methods to interpolate the values across the entire nation. After kriging, we calculate an additional weather variable, daily range in temperature. To aggregate the daily data to the weekly unit, we compute summary statistics (minimum, mean, and maximum) of each value over the seven days. Similar to the satellite variables, the weekly weather variables are then aggregated to the state level by computing summary statistics (minimum, mean, maximum, and standard deviation) over the municipalities. This results in a total of 40 weather variables for each state.

\begin{figure}[ht]
\includegraphics[width=1\textwidth]{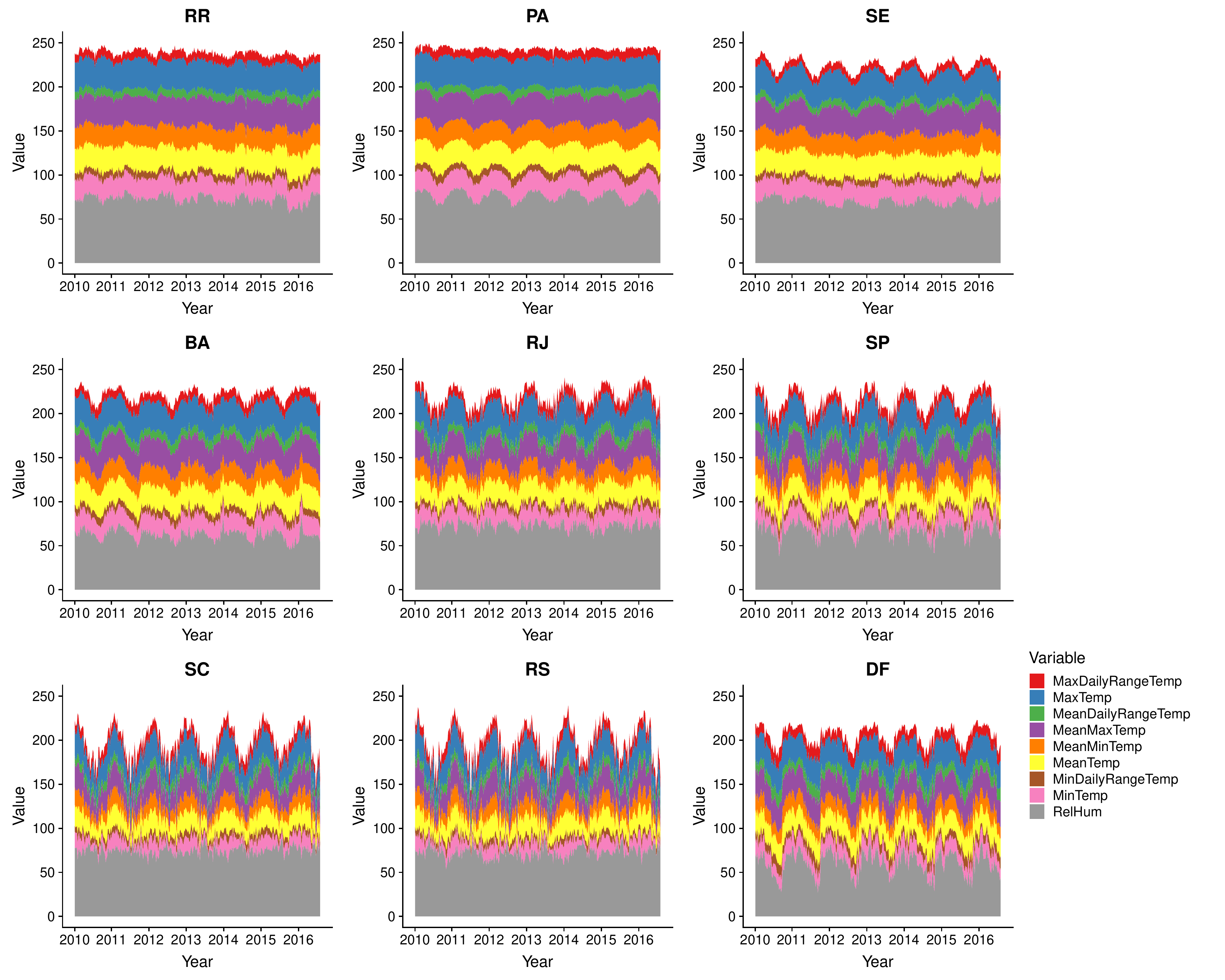}
\centering
\caption{\textbf{Weather variables in each Brazilian state.} For a selection of Brazilian states, weather variables are plotted from 2010 to 2016. (Variables with approximately zero variance in all states are omitted.) Plots for the remaining 18 states are included as supplementary materials.} 
\label{fig:weather}
\end{figure}

For each state, the weather variables corresponding to the mean summary statistic (over the municipalities) are plotted from 2010-16 in Figure \ref{fig:weather}. As expected, these variables are strongly periodic, peaking at approximately the same time each year. However, there are notable differences between states in terms of volatility and the amplitude of seasonality. For example, the states RJ, SP, SC, RS, and DF are the most volatile here, and PA is the least. DF tends to exhibit a higher seasonal amplitude in its variables (reflecting more extreme seasonal changes in weather), while states like RR have lower seasonal amplitudes. Such spatial differences reinforce the importance of modeling at the state, rather than the national level. Meanwhile, the prominent seasonality of these variables inspire consideration of time series methods that can handle such data, e.g., seasonal differencing or seasonal trend decomposition based on LOESS (STL). 

%_______________________________________________________________________________________________
%_______________________________________      Old text    ______________________________________
%_______________________________________________________________________________________________

\begin{comment}
%The association between climate (e.g., temperature and precipitation) and dengue risk has been reported in the literature. For example, a meta-analysis by Fan et al. \cite{fan2014systematic} showed a positive and increased association for temperature between $22\,^{\circ}{\rm C}$ and $29\,^{\circ}{\rm C}$. 

%%\Jnote{Provide Supplementary Methods on kriging?}

\end{comment}

\subsubsection*{Google Search Trends}
We have collected weekly data from Google Health Trends, since Google is the most widely used search engine and has a 97\% market share in Brazil \cite{statcounterBR}.
Google provides de-identified, normalized, region-specific trends in search activity via the Google Trends website \cite{google}. At the country and state levels of Brazil from January 2011 to June 2017, we have obtained data 
for 19 dengue-related keywords (listed in Table \ref{table:data}), including Portuguese and English terms for ``dengue", ``mosquito," and ``aedes aegypti," the primary vector of dengue \cite{who_2015}.
%However, the keyword ``DF" was omitted from analysis, due to confounding with the Distrito Federal unit of Brazil. 

\begin{figure}[ht]
\includegraphics[width=1\textwidth]{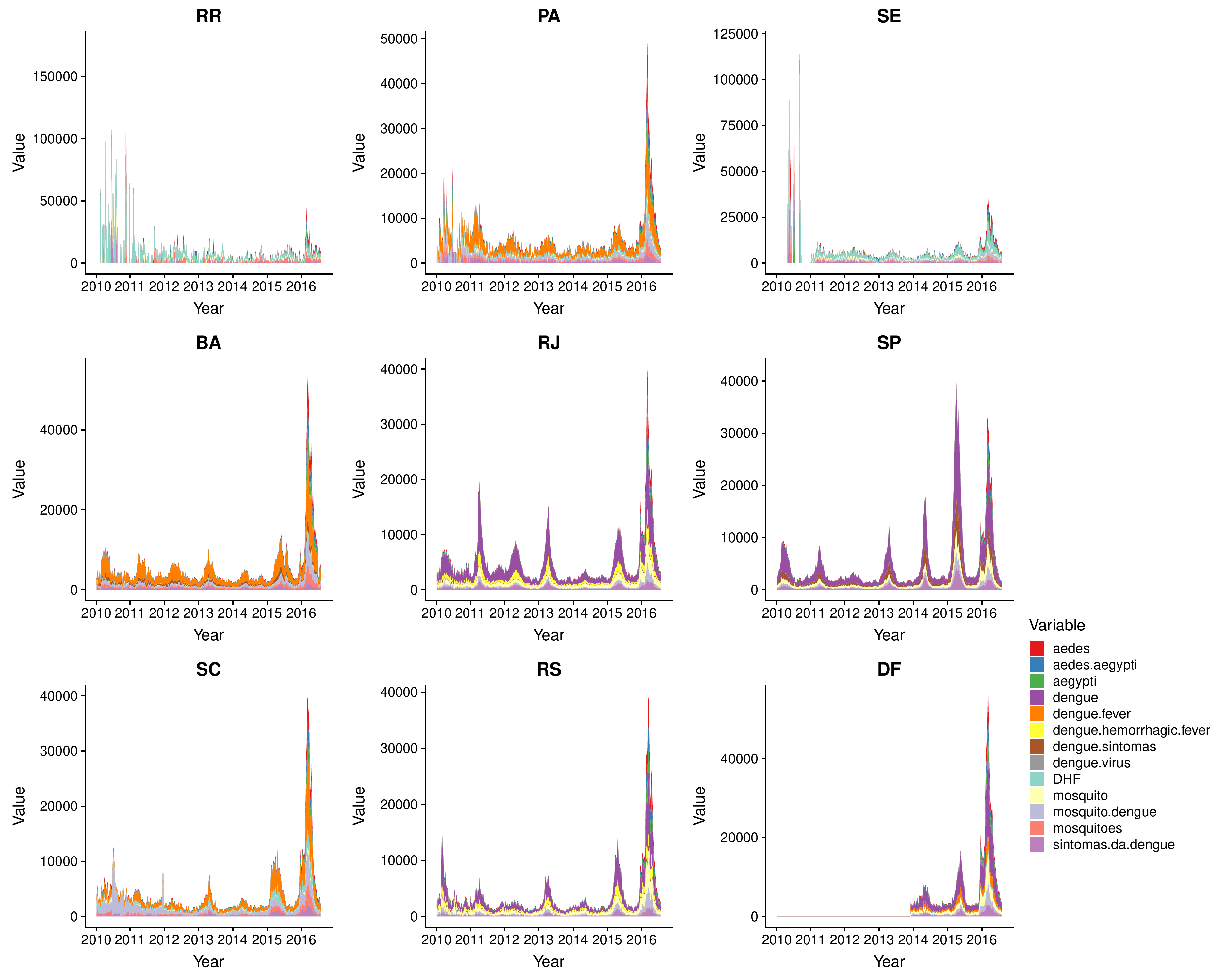}
\centering
\caption{\textbf{Google variables in each Brazilian state.} For a selection of Brazilian states, Google Health Trends variables are plotted from 2010 to 2016. (Variables with approximately zero variance in all states are omitted.) Plots for the remaining 18 states are included as supplementary materials.} 
\label{fig:GHT}
\end{figure}

In Figure \ref{fig:GHT}, the values of the Google variables are shown for a selection of states over the years 2010-16 (DF has Google data unavailable prior to 2014). For most displayed variables, there are seasonal spikes approximately coinciding with dengue season. However, the variables differ substantially across space and time. For many states, the variables are more volatile in 2010-11 than in 2012-16, and SE features gaps in 2010-11. The state RR exhibits very high noise in its Google variables, compared to the more periodic seasonal patterns formed in the other states. Among most states (PA, BA, RJ, SC, RS, and DF), a prominent global maximum is attained in 2016, while SP peaks in 2015 (but also attains a very high local maximmum in 2016), and SE peaks in 2010-11. %but the other local maxima tend to differ in relative height; e.g, RJ's second highest seasonal peak occurs in , while PA's occurs in 2015. 2011
Perhaps most notably, the most common Google search keywords differ substantially by state. Searches for ``dengue fever" are the most common in PA, BA, and SC; searches for ``dengue", ``sintomas da dengue", ``mosquito", and ``dengue hemorrhagic fever" are dominant in RJ, SP, RS, and DF; and searches for ``DHF" are popular in RR and SE. Among the considered exogenous data streams (out of Google, satellite, and weather), the Google variables feature the greatest distinctions temporally and spatially.

%The states RR, SC, and DF exhibit high noise in their Google variables, while the remaining states PA, SE, BA, RJ, SP, and RS tend to form more periodic seasonal patterns. Among the latter group of states, a prominent global maximum is attained in 2016, but the other local maxima tend to differ in relative height; e.g, RJ's second highest seasonal peak occurs in 2011, while PA's occurs in 2015. Perhaps most notably, the most common Google search keywords differ substantially by state. Searches for ``DHF" are the most common in RR, SC, and DF; searches for ``dengue fever" are dominant in PA, SE, BA, and SP; and searches for ``dengue" and ``dengue hemorrhagic fever" are popular in RJ and RS. Among the considered data streams (out of Google, satellite, and weather), the Google variables feature the greatest distinctions temporally and spatially. %Hence, the Google data are expected to be particularly helpful for predicting dengue at the state level of Brazil. 

\subsubsection*{Dengue Case Data}
The Brazilian Ministry of Health for Dengue provided the epidemiological dengue data from January 3, 2010 to July 17, 2016. This dataset contains newly diagnosed cases per epidemiological week for each of the 5564 municipalities. In Brazil, dengue case investigation forms include information on basic demographic data, dates of symptom onset and sample collection, case classification (dengue fever, DHF, DSS, or discarded case), and outcome. Individual data are locally entered into Brazil's National Reportable Disease Information System (SINAN) and subsequently transmitted to state and national levels \cite{siqueira2005dengue}. These data are considered “gold standard” for mosquito-borne disease reporting, covering both time and space in Brazil. However, like most disease surveillance data, they are subject to underreporting; SINAN contains only about half the cases reported to the National Unified Health System (SUS) \cite{coelho2016sensitivity}.

\begin{figure}[htpb]
\includegraphics[width=.8\textwidth, trim={0cm 3cm 0cm 2cm},clip]{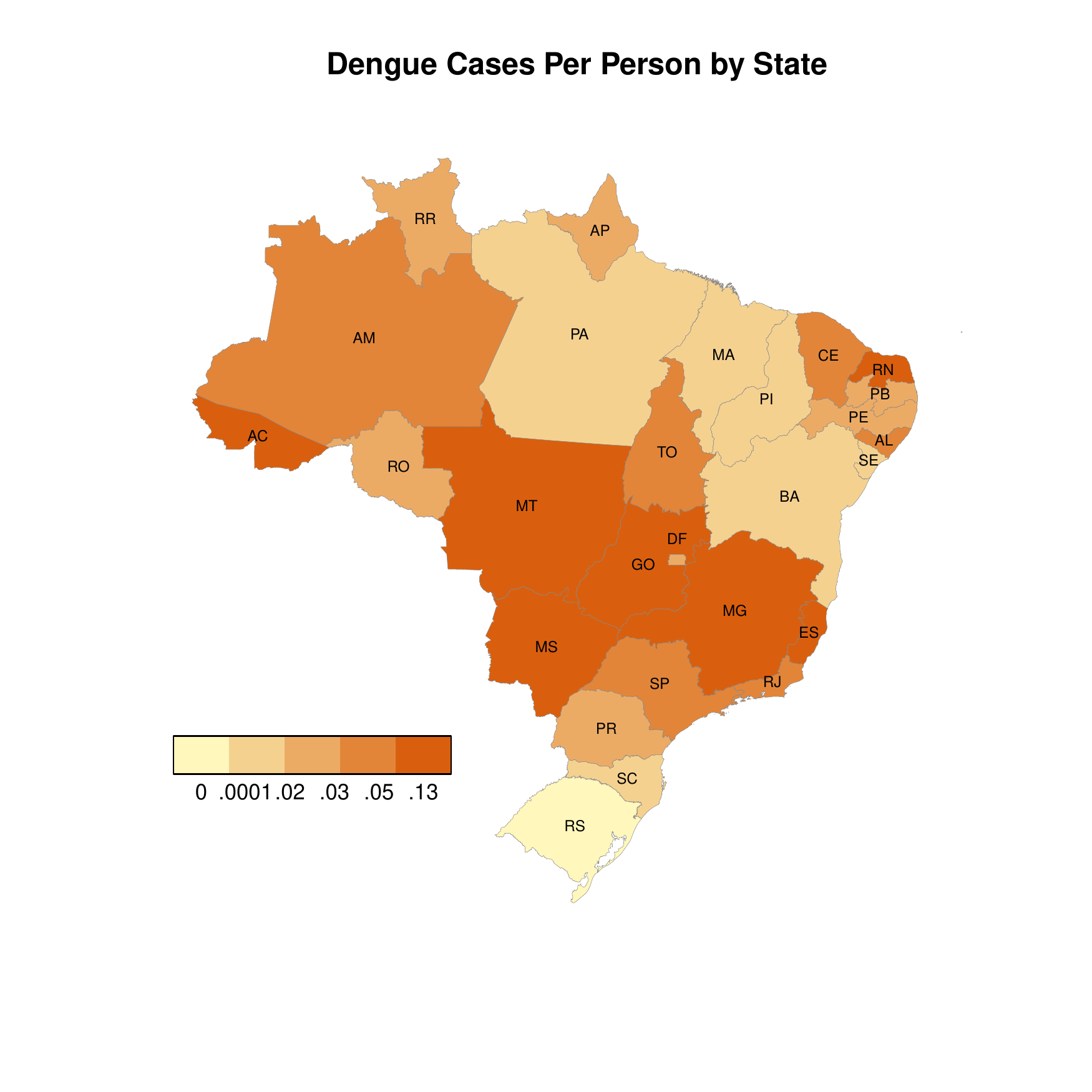}
\centering
\caption{\textbf{Number of dengue clinical cases per person in each Brazilian state.} For each state, the number of dengue clinical cases (from January, 2010 to July, 2016) per person is displayed. Map breaks are chosen as quartiles.} 
\label{fig:cases_per_person}
\end{figure}

\begin{figure}[htpb]{}
\includegraphics[width=1\textwidth]{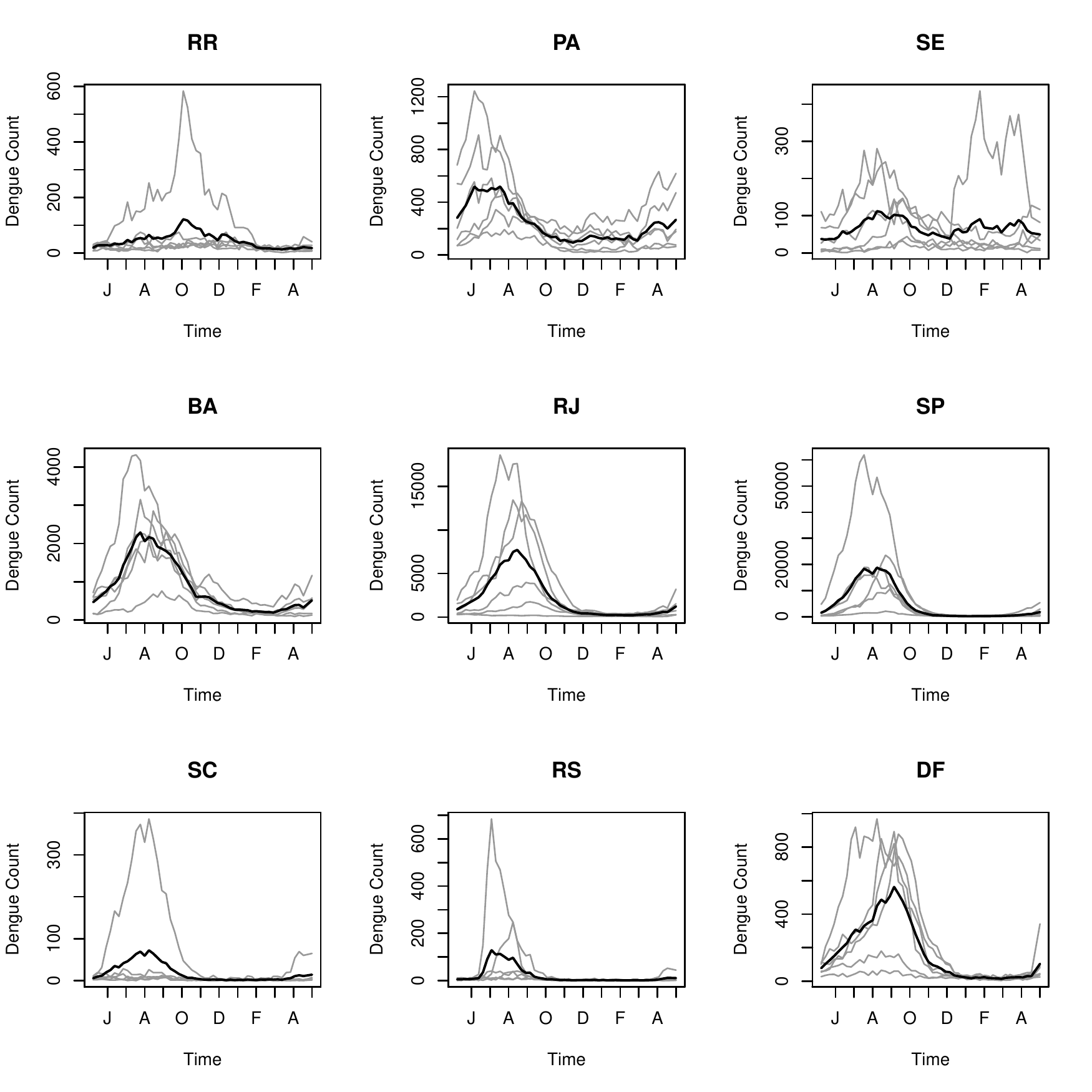}
\centering
\caption{\textbf{Mean dengue cases over epidemic weeks in each Brazilian state.} For a selection of Brazilian states, the number of dengue cases per epidemic week (computed over all weeks from 2010 to 2015) is displayed in gray from June to May. The black lines correspond to the mean number of dengue cases per epidemic week during the same period. The year 2016 is omitted due to unavailable data for the full year. The remaining 18 Brazilian states' case counts are plotted in the supplementary materials section.} 
\label{fig:cases_avg}
\end{figure}

The dengue cases are summarized spatially and temporally in Figures \ref{fig:cases_per_person} and \ref{fig:cases_avg}, respectively. Figure \ref{fig:cases_per_person} maps the total number of dengue cases per person for each state, where population estimates are obtained from the 2010 Brazilian state census data. Among the states, cases range from approximately 0.0923\% to 13.2\% of the population, with states in the Center-West (DF, GO, MS, MT) and Southeast (ES, MG, RJ, SP) regions experiencing the greatest burden. In Figure \ref{fig:cases_avg}, the number of dengue cases from 2010 to 2015 is displayed for each epidemic week in a state. As expected, these curves generally exhibit symmetric epidemic curves in the June-October months (``dengue season"), peaks around April, decreases until December, and small increases starting April. Surprisingly, this pattern is less pronounced for some states: RR and DF experience peaks as late as September-October, while PA and SE have approximately bimodal structures. There are several potential explanations for these results: 1) some states' dengue transmission may deviate from the classic dengue epidemic curve, e.g., due to presence of new serotypes or tourism effects; 2) some states' clinical case data contain reporting biases, e.g., misreporting of chikungunya or Zika cases as dengue; or 3) small states experience greater sampling variation. Regardless, this spatial variation in dengue cases encourages a model-fitting and nowcasting process specific to each state.

%_______________________________________________________________________________________________
%_______________________________________      Old text    ______________________________________
%_______________________________________________________________________________________________

\begin{comment}

%However, it is important to note that this data is subject to underreporting. Of cases reported to National Unified Health System (SUS), only about half were also reported to Brazil's National Reportable Disease Information System (SINAN)\cite{coelho2016sensitivity}, where our data is derived from.

\end{comment}
\subsection*{Predictive Modeling Approach}

%_______________________________________________________________________________________________
%______________________________ Notation for next sections  ______________________________________
%_______________________________________________________________________________________________
\begin{comment}
\textcolor{red}{Notation}
\begin{enumerate}
    \item $\{y_t\}$ is for dengue count at time $t$
    \item each week is denoted by $t=1,\hdots,n$
    \item diverse data streams are $\{z_{t1}\}$, \hdots, $\{z_{tm}\}$
    \item $p,d,q,P,D,Q$ are ARIMA orders
    \item $i,j,k$ are dummy variables 
    \item $l$ denotes lags outside of the context of autoregressive or MA models
    \item if referring to the number of variables (for a subset) $k \leq m$ is the preferred dummy variable 
    \item $\{u_t\},\{w_t\}$ denote white noise time series 
\end{enumerate}
\end{comment}

%_______________________________________________________________________________________________

Literature review (Introduction) and exploratory analysis (Methods - Data) have suggested the importance of diverse data streams for accurately modeling dengue. Each of the variables in our dataset (out of Google, weather, satellite, and clinical dengue case count) generally feature distinctions across states and time. Meanwhile, obtaining specialized forecasts for each state in Brazil might benefit public health decision-making; by understanding the risks specific to each area, specialists can more effectively tailor resources and policy. Hence, we develop the following spatiotemporal modeling framework: dengue is separately modeled for each of the 27 states and assumed to be affected only by neighboring states. For example, the state of Rio de Janeiro (RJ) can be modeled using satellite, weather, and Google data defined for the state, as well as dengue case counts for the states SP, MG, and ES; forecasts are subsequently generated specific to RJ. Unlike most past work on dengue modeling, we do not assume there is one ``best" model across all states and instead allow each state to potentially be modeled differently than others. While spatial interactions between states may be more complex, we believe our framework to be a good balance between model simplicity, nowcast accuracy, and practical application to public health. All predictive modeling is implemented in the statistical programming environment R \cite{R}, and task parallelism (fitting and nowcasting across the 27 states) is performed through the R packages \textit{snow}\cite{snow} and \textit{parallel}\cite{parallel}.

\subsubsection*{Cross-Validation}
Due to limited availability of Brazilian case count data, we estimate the predictive performance of our methods within the dataset.
%To measure the nowcast error, we would ideally compare true dengue counts to the nowcasted dengue counts to obtain ``true" error. However, access to official Brazilian case counts is limited, so our most recent dengue counts are from July, 2016. Until we receive more recent data, we must instead assess nowcast error within the given dataset. 
Such estimation is prone to optimistic results from model ``over-fitting", i.e. fitting the model to the specific dataset rather than to the underlying data-generating phenomenon. Cross-validation (CV) is essential to provide less biased estimates of model error (or accuracy), so we implement CV through independent training and testing weeks. There are 342 weeks of dengue data total: we assign weeks 1 through 261 to be for training, and weeks 262 through 342 for testing. This approximate $76:24$ training:testing split of weeks is chosen to ensure at least one full year for testing, while having sufficient training data (five full seasons) to account for high variation in dengue count between years. 
%the five full dengue seasons for training and approximately two for testing. 
We choose as metrics the root mean square error (RMSE), relative RMSE (RRSME), Pearson correlation, mean absolute error (MAE), and relative MAE (RMAE), defined here as
\begin{equation}
\begin{aligned}
\textrm{RMSE}&=\sqrt{\frac{1}{n}\sum_t (y_t-\hat{y_t})^2}, \\
\textrm{RRMSE}&=\frac{\sqrt{\frac{1}{n}\sum_t (y_t-\hat{y_t})^2}}{\sum_t y_t}=\frac{\textrm{RMSE}}{\sum_t y_t}, \\
R&=\frac{\sum_t (y_t - \bar{y}) (\hat{y_t} - \bar{\hat{y_t}}) }{\sqrt{\sum_t (y_t - \bar{y})^2 } \sqrt{\sum_t (\hat{y_t} - \bar{\hat{y}})^2 } }, \\
\textrm{MAE}&=\frac{\sum_t |y_t - \hat{y_t}|}{n}, \textrm{ and} \\
\textrm{RMAE}&=\frac{\textrm{MAE}}{\sum_t y_t},
\end{aligned}
\end{equation}
where $y_t$, $\hat{y_t}$, $\bar{y_t}$, and $\bar{\hat{y_t}}$ are, respectively, observed dengue, predicted dengue, sample mean dengue, and sample mean predicted dengue within a state for week $t$ out of $n$ total weeks. The predictive performance must be assessed from the testing set, so we set the summation indices to $t =262,\hdots,342$ when calculating these metrics. RMSE and MAE are some of the most common performance metrics for quantitative data and can be used to compare model performance, but they are not unitless measures and hence cannot be compared across states. RRMSE, $R$, and RMAE are unitless alternatives that can be used to compare performance across states.  

To simulate real-time dengue prediction in Brazil, we obtain nowcasts for the testing weeks in each state using only data we would expect to be available at that time. In practice, we expect that health agencies can provide dengue case reports within two weeks of incidence. Hence, for every testing week, we update our training dengue case count variables from as recent as two weeks prior and produce 2-step-ahead forecasts (which are thus nowcasts corresponding to current time $t$). More precisely, for week $t$ in the testing set (where $t$ is an integer above 4), we train a model using all 261 training weeks plus the prior testing weeks, $262,263,\hdots,262+t-4$. When neighboring dengue case counts are included as predictors in a model for a state, the two week lag time is addressed by forecasting the values for the two weeks through a simple SARIMA model. Meanwhile, the weather, satellite, and Google variables can be updated in real-time, so we allow dengue to be forecasted at time $t$ with values of these variables for as recent as time $t$.

\subsubsection*{Dimension Reduction of Exogenous Variables}
Several of our models presented in the next section include the satellite, climate, Google Health Trends variables, and dengue case counts from neighboring states. Unless otherwise noted, we refer to these variables as exogenous variables and denote them as $\{z_{t1}\},\{z_{t2}\},\hdots,\{z_{tm}\}$, while the true dengue count for the modeled state is considered the endogenous variable and denoted as $\{y_t\}$. Here, we discuss our motivations and methods for reducing their dimension.

There are up to $m=115$ exogenous variables per state, and including all or most of those variables in our models could violate model assumptions and decrease nowcast accuracy. The models in the next section require the exogenous variables to be uncorrelated, which is an assumption frequently violated; for example, four summary statistics (minimum, mean, maximum, and standard deviation) of the maximum temperature are represented as variables and expected to have high correlations among themselves. %Additionally, many variables have high cross-correlation because of seasonal relationships; e.g., the Google Health variable “mosquito” tends to increase during January, during which temperature also peaks.
Additionally, including too many predictors can contribute to over-fitting, resulting in severe reductions of testing accuracy. 
To address such issues, we apply techniques of dimension reduction, expressing information from the full set of variables in a smaller, more concise subset.

First, we omit any variables with zero-variance in the training or testing set. Then we consider principal component analysis (PCA) and partial least squares (PLS), two well-established methods from statistics that transform and reduce the dimension of the exogenous variables through singular value decomposition (SVD). PCA and its variants have been applied to time series analysis \cite{shang2014survey}. Briefly, PCA is an orthogonal linear transformation that maximizes the covariance of the predictors in the lowest dimension possible (through the top “principal components”). PCA is considered an unsupervised method, because it does not incorporate information about the response variable. %Hence, it is possible that the top ``principal components” are not correlated with the response, rendering them unrelated to dengue forecasting. To avoid such a situation, we identify the five principal components that are most strongly (Pearson) correlated with dengue count. 
PLS can be viewed as a supervised modification of PCA, transforming the variables such that they are strongly correlated with the response, uncorrelated among themselves, and maximize covariance.
The relationship between these methods is discussed in \cite{dunn1989principal}, among other works. Our approach for incorporating principal components and partial least squares scores into a model is specific to the model, so we refer further explanation to the next section.

\subsection*{Individual Predictive Models}
We discuss the individual models considered for nowcasting dengue in Brazil. For a fixed state, all models are separately fit for that state, using the expanding training weeks as previously defined. Subsequently for that state, recursive nowcasts (2-step ahead forecasts) are obtained for the testing weeks. 

We consider six different time series models, chosen because they are well-understood in statistics, explicitly account for temporal dependency, and range in complexity and inclusion of variables. These models, as well as two methods for combining them (detailed in a later section), are summarized in Table \ref{table:model_summary}. SARIMA (1) and the STL models (4, 5) contain only past dengue case counts, while SARIMAX (2, 3) and VAR (6) additionally contain the exogenous variables from heterogenous data streams (Google Health trends, weather, satellite, and neighboring state case counts). Hence, models (1, 4, 5) can represent nontrivial baselines that can be compared against the models combining novel and traditional data streams (2, 3, 6). For each model, we produce both point estimates and the standard 95\% prediction intervals based on asymptotic normality of errors. We note that the dengue response is not constrained to be a count, which would limit choice of time series models and availability of statistical software. In some rare cases, dengue forecasts are slightly below 0, which we simply convert to 0. %The SARIMAX (2, 3) nowcast dengue based on past dengue counts, as well as the exogenous variables (Google Health trends, weather, and satellite) that have been transformed through singular value decomposition. VAR (5) is a multivariate model containing past dengue and other variables from the data streams as a response vector. 

\begin{table}[ht]{}
    \centering
    \begin{tabular}{cll}
    \hline
        \textbf{Type} &\textbf{Model} &\textbf{Description} \\
        \hline
 \textbf{Individual Models} &(1) Seasonal autoregressive integrated moving average (SARIMA) & no exogenous variables\\
            &(2) SARIMAX with principal component analysis (PCA) & exogenous variables\\
             &(3) SARIMAX with partial least squares (PLS) & exogenous variables\\
             &(4) Additive seasonal trend decomposition based on LOESS (STL) & no exogenous variables  \\
             &(5) Multiplicative STL & no exogenous variables  \\
             &(6) Vector autoregression with PCA-transformed data & ``exogenous"$^1$ variables\\
            \hline
\textbf{Combining Models} &(7) Trimmed mean ensemble &robustly computes mean of predictions\\
                &(8) Weighted mean ensemble &computes weighted mean of predictions\\
        \hline
    \end{tabular}
    \caption{\textbf{Summary of models to nowcast dengue in each Brazilian state.} We list the individual models of dengue and whether they include exogenous variables or no exogenous variables. The exogenous variables include Google, weather, and satellite data (from as recent as the current week), as well as neighboring state case counts (from as recent as two weeks prior). All individual models additionally contain past dengue count for the state (from as recent as two weeks prior). We also list ensemble methods, which aggregate nowcasts from the individual models. 
    $^1$``Exogenous" variables are technically considered endogenous in a VAR model, which is multivariate.
    %$^2$Model selection fits all individual models in training, selects the one minimizing training RMSE, and applies that model to testing. $^3$Ensemble fits a linear regression model of dengue count on forecasts in training, then predicts the testing weeks from that model.
    }
    \label{table:model_summary}
\end{table}

\subsubsection*{Seasonal Autoregressive Integrated Moving Average Model}
SARIMA is one of the most popular models in time series analysis. The model can be denoted by ARIMA$(p,d,q)\times (P,D,Q)$, where $p,d,q$ are respectively the autoregressive, difference, and moving average components and $P,D,Q$ are the seasonal autoregressive, difference, and moving average components. Define $B$ to be the backshift operator, i.e., %letting $\{y_t\}$ be the dengue count time series, 
$By_t=y_{t-1}$. We can define order $k$ of $B$ as $B^ky_t=y_{t-k}$, where $k$ is a nonnegative integer. Similarly, define $\Delta$ as the difference operator, i.e., $\Delta^k=(1-B)^k$ for any nonnegative integer $k$. The model can be expressed as
\begin{equation}
\label{eq:SARIMA}
    \Phi_P (B^S) \phi(B) \Delta_S^D \Delta^d y_t=\delta + \Theta_Q(B^S)\theta(B)w_t,
\end{equation}
where $\delta$ is an intercept, $w_t$ is white noise, $\phi(B)$ and $\theta(B)$ are respectively autoregressive and moving average components of orders $p$ and $q$ (i.e., $\phi(B)=1-\phi_1B-\phi_2B^2-\hdots-\phi_pB^p$ and $\theta(B)=1+\theta_1B+\theta_2B^2+\hdots,\theta_qB^q$), $S$ is the seasonal order, $\Phi_P(B^S)$ and $\Theta_Q(B^S)$ are seasonal autoregressive and moving average components with orders $P$ and $Q$, and $\Delta^d=(1-B)^d$ and $\Delta_S^D=(1-B^S)^D$ are the ordinary and seasonal difference components \cite{shumway2017time}. In practice, a major challenge of forecasting through SARIMA is choosing appropriate orders, which can drastically affect predictive performance. Additionally, a fundamental assumption of SARIMA is that the time series $\{y_t\}$ is stationary.

Any nonstationarity is addressed by first performing the Box-Cox transformation $f_\lambda:\{y_t\} \rightarrow \mathbb{R}$ defined by
\begin{equation}
    f_\lambda(y_t) =
  \begin{cases}
   \frac{{y_t}^\lambda - 1}{\lambda}, & \text{if } \lambda \neq 0 \\
    log(y_t), & \text{if } \lambda = 0,
  \end{cases}
\end{equation}
where the parameter $\lambda$ is estimated from the time series $\{y_t\}$.  
Next, the difference orders $d$ and $D$ are constrained to be less than 3 and 2, respectively, and chosen through unit root tests. The remaining orders $p,q,P$, and $Q$ are identified through model selection. To reduce the number of possible combinations and the potential for overspecified models, we consider only the orders $p=0,1,\hdots,5$, $q=0,1,\hdots,5$, $P=0,1,2$, and $Q=0,1,2$. For each combination of these orders, we fit a model and assess the fit through some chosen criterion. We choose a popular criterion from statistics, Akaike’s Information Criterion (AIC), defined in general as 
\begin{equation}
\label{AIC}
    \textrm{AIC} =-2 log(L) + 2K,
\end{equation}
where $L$ is the maximized likelihood of the model fitted to the data and $K$ is the number of parameters in the model \cite{ akaike1998information}. For SARIMA specifically, $K=p + q + P + Q + o$, where $o=1$ if the intercept $\delta \neq 0$ and $0$ otherwise \cite{hyndman2007automatic}. (We tested another well-established criterion, the Bayesian Information Criterion (BIC) \cite{schwarz1978estimating}, but it often produced models that were too sparse.) Here, we determine the ``best" model for a state to be the one whose orders yield the lowest AIC. This modeling process is implemented through the R package \textit{forecast} \cite{hyndman2019package}.

\subsubsection*{SARIMAX}
For methods (2) and (3), we incorporate the exogenous variables from diverse data streams by generalizing the SARIMA model from (1). The exogenous variables are transformed through PCA in (2) and through PLS in (3).

%Let $\{z_{t1}\},\{z_{t2}\},\hdots,\{z_{tm}\}$ be the $m$ time series variables from diverse data streams, and
Let $\{s_{t1}\},\{s_{t2}\},\hdots,\{s_{tk}\}$ be the $k$ transformed, dimension-reduced variables from PCA or PLS, where $k$ is an integer satisfying $k \leq m$. Then the model for (2) or (3) can be denoted by SARIMAX$(p,d,q) \times (P,D,Q)$, defined as
\begin{equation}
\label{eq:SARIMAx}
y_t=\beta_1 s_{t1}+\hdots + \beta_k s_{tk} + u_t,
\end{equation}
where $\beta_j$ for $j=1,\hdots,k$ is a coefficient, and the error $u_t$ is modeled as SARIMA$(p,d,q)\times (P,D,Q)$, i.e.,
\begin{equation}
     \Phi_P (B^S) \phi(B) \Delta_S^D \Delta^d u_t=\delta + \Theta_Q(B^S)\theta(B)w_t,
\end{equation}
with all notation as in Equation \ref{eq:SARIMA}. 

A similar model selection process is followed for PCA and PLS. We retain a dimension of five to encourage model parsimony and reduce risk of over-fitting. For each of the $5^2=32$ nonempty subsets of transformed variables, we fit a SARIMAX model (with model orders simultaneously identified following the same AIC-minimizing process as above). The final model is chosen as the subset of variables and combination of model orders with minimum AIC. The distinction is that for PLS, we reduce the dimension to five by choosing the top partial least squares scores. For PCA, we choose the five principal components most strongly correlated with dengue case count. Since PCA is unsupervised, without the correlation condition, it is possible that the top ``principal components" could be unrelated to dengue forecasting.

\subsubsection*{Seasonal Trend Decomposition Based on LOESS (STL)}
Seasonal trend decomposition based on LOESS (STL) is a nonparametric, univariate decomposition of a time series into three components: seasonal, trend, and remainder \cite{cleveland1990stl}. STL is among the most popular time series decomposition methods, and its advantages over others include its fast computation, flexibility in extent of trend or seasonality smoothing, robustness to outliers, and greater specification of the seasonal component's period \cite{cleveland1990stl}. 
We have considered both additive and multiplicative forms of the model, i.e.,
\begin{equation}
\label{eq:STL_add}
    y_t=T_t + S_t + R_t
\end{equation}
and 
\begin{equation}
\label{eq:STL_mult}
    y_t=T_tS_tR_t,
\end{equation} 
respectively, where $T_t$, $S_t$, and $R_t$ are the trend, seasonal, and remainder components at time $t$. The components $T_t$ and $S_t$ are estimated through successive passes of a nested (inner and outer) loop algorithm, which applies several smoothers (LOESS \cite{cleveland1979robust} and moving average) and robustness weights to flexibly and robustly fit the data. There are six main parameters that must be specified: the periodicity ($D$) of the time series, the number of inner and outer loops ($i, o$, respectively), and the span of the LOESS window for seasonal component estimation, trend component estimation, and the low-pass filter ($n_s, n_t, n_l$, respectively). We specify the seasonal periodicity as $D=52$, to reflect the yearly pattern followed by weekly dengue count. The respective parameters $i$ and $o$ are chosen as 1 and 15, the recommended values for robust fitting of STL \cite{hyndman2019package}. The low-pass filter span $n_l$ is specified as 53, the minimum odd integer greater than or equal to $D$, as suggested in \cite{hyndman2019package}. We have specified the remaining parameters as $n_s=155$ and $t_s=25$, by comparing model fit within the training set for a range of odd values.

The seasonal component $S_t$ is forecasted through an exponential state smoothing model     \cite{hyndman2008forecasting}, which visually produces forecasts repeating a similar periodic pattern. The deseasonalized time series ($Y_t - S_t$ for an additive model and $\frac{Y_t}{S_t}$ for multiplicative) is forecasted through SARIMA, following the same process outlined in the SARIMA section. Finally, the forecasted series is reconstructed by summing or multiplying the predicted components, similarly as in Equations \ref{eq:STL_add} and \ref{eq:STL_mult}. The SARIMAX fitting and forecasting is implemented through the R package \textit{forecast}.

\subsubsection*{Vector Autoregression (VAR)}
For each time step $t$, the SARIMAX model in Equation \ref{eq:SARIMAx} includes the exogenous variables only at $t$. However, these variables may relate to the dengue count $\{y_t\}$ at lagged times. To accommodate such lags, we consider the vector autoregression model (VAR).

VAR is a popular, flexible, and powerful method for modeling and forecasting multiple time series \cite{lutkepohl2005new}. In contrast to the previously presented models, VAR is multivariate, modeling multiple response variables simultaneously as a response vector. While our priority is forecasting dengue (not other variables), a multivariate model can help capture complex temporal dynamics between variables and improve the accuracy of dengue forecasts.

Let $\textbf{y}_t=(y_{t1},y_{t2},\hdots, y_{tk})^T$ denote a $k \times 1$ response vector at time $t$, where $y_{t1}=y_t$ is the dengue count for a state and for each $i=2,\hdots,k$, $y_{ti}=z_{ti}$ is one of the $m$ ``exogenous" variables. (For clarity, we continue referring to these variables as ``exogenous", even though they are considered endogenous in a multivariate model such as VAR.) A VAR($p$) model can be specified as
\begin{equation}
\label{eq:VAR}
    \textbf{y}_t=\textbf{A}^1 \textbf{y}_{t-1}+\hdots+\textbf{A}^p\textbf{y}_{t-p}+\textbf{w}_t,
\end{equation}
where $\textbf{A}^i$ for $i=1,\hdots,p$ are $k \times k$ coefficient matrices and $\textbf{w}_t$ is a white noise vector \cite{lutkepohl2005new}. 

With some algebra, each entry of the response vector $\textbf{y}_t$ can be expanded from Equation \ref{eq:VAR}, e.g., dengue count can be expressed as 
\begin{equation}
\label{eq:VAR_expand}
    y_{t1}=y_t=A_{11}^1 y_{t-1,1}+\hdots + A_{1k}^1 y_{t-1,k}+
                A_{11}^2 y_{t-2,1} + \hdots + A_{1k}^2 y_{t-2,k}+
                \hdots +
                A_{11}^p y_{t-p,1} + \hdots + A_{1k}^p y_{t-p,k}
                + w_{t1},
\end{equation}
where $A_{MN}^i$ is the entry in row $M$ and column $N$ in the coefficient matrix $\textbf{A}^i$, for $i=1,\hdots,p$. This equation for dengue will be used to obtain forecasts in the testing weeks. Similarly to the models from previous sections, a challenge of VAR in practice is selecting appropriate variables and the autoregressive order $p$. To reduce model complexity and risk of over-fitting, we limit the maximum order to be $p=8$.

We apply PCA to the set of ``exogenous" variables. (PLS is not applied here, as its supervised algorithm would fail to identify relationships between lagged variables.) Similarly as in SARIMAX with PCA, we seek the principal components that might help predict dengue. We calculate the cross-correlation function between dengue count and each principal component for lags up to 8 (lags beyond 8 need not be considered, since we have constrained the autoregressive order $p$ to be less than or equal to 8); the principal components are ranked by strength of cross-correlation when leading dengue and the top five are selected. %reducing the space of potential variables to under 6 (dengue plus up to 5 ``exogenous" variables). 
To identify an appropriate autoregressive order $p$ for the VAR($p$) model, we proceed with model selection on this reduced feature set. We fit a model in training for all combinations of the five principal components and autoregressive orders $p=1,2,\hdots,8$. The ``best" model is selected as the one that minimizes the AIC \cite{akaike1998information}, chosen for similar reasons as for SARIMA and defined in Equation \ref{AIC} with $K=p(k+1)+2$ (where $k$ is the length of the response vector). 
%(We also tested the Bayesian Information Criterion (BIC)\cite{schwarz1978estimating}, but it yielded higher testing forecast error on average.) 
This model is subsequently used to forecast within the testing weeks.
The VAR fitting and forecasting 
is implemented through the \textit{vars} package in R \cite{pfaff2008var}.

\subsection*{Combining Predictive Models}
As demonstrated in the Results section, the ``best" statistical model of dengue (achieving minimum testing error) differs across states. However, high temporal variation in dengue count indicates that a state's ``best" method in the testing weeks (2015-16) might not be the best in the future. For real-time, practical nowcasting, we consider ensemble methods to automatically generate robust predictions for each state. Ensembles are statistical learning models that combine individual models, with the goal of capturing the strengths of each individual in a diverse set \cite{dietterich2000ensemble, hastie2005elements}.
Theoretically and empirically, ensembles have often outperformed individual models for general prediction tasks \cite{dietterich2000ensemble}, as well as time series prediction \cite{allende2017ensemble}. Common approaches involve perturbing the data (e.g., through bagging), altering the individual models (e.g., by adding regularization terms), or aggregating the outputs from individual models (e.g., by computing the mean)\cite{allende2017ensemble}. Due to the good performance of our individual models in training, we expect simple, common aggregation methods to perform well and consider a trimmed and weighted mean.

A trimmed mean robustly measures center by removing extreme observations from a sample before computing its mean \cite{allende2017ensemble}. We specify the trimmed mean to remove the most extreme 20\% of observations, which here corresponds to omitting the minimum and maximum nowcasts. For each state and week $t=262, \hdots, 342$ in testing, the trimmed mean ensemble nowcast is computed as the mean of the trimmed sample $\{\hat{y}_{t_{(2)}}, \hat{y}_{t_{(3)}}, \hdots, \hat{y}_{t_{(5)}}\}$, which has omitted the minimum and maximum nowcasts $\hat{y}_{t_{(1)}}$ and $\hat{y}_{t_{(6)}}$. To produce 95\% prediction intervals, we use a conservative approach: the bounds are selected as the minimum and maximum of the 95\% prediction interval bounds (previously computed through the standard asymptotic normality procedures) associated with the nowcasts in the trimmed sample $\{\hat{y}_{t_{(2)}}, \hat{y}_{t_{(3)}}, \hdots, \hat{y}_{t_{(5)}}\}$.

Weighted means are another common approach in aggregation-based ensembles, where the weights are often chosen based on individual model performance in the training or validation set \cite{allende2017ensemble}. For the purposes of selecting weights, we divide our main training set into a smaller training set (weeks $t=1, 2, \hdots, 157$) and a validation set (weeks $t=158, 159, \hdots, 261$). On the validation set, we refit each individual model by mirroring the previously discussed cross-validation approach. The weights for a state are selected as the proportion of times an individual model's prediction has minimized the $l_1$ norm (between observed and predicted dengue) in the validation set. For example, suppose that $\hat{y_t}=\hat{y_t}^{\textrm{VAR}}$ minimizes $|\hat{y_t} - y_t|$ only for the three weeks $t=158, 159, 162$; then the weight for VAR would be $\frac{3}{261-158+1}$. For each state, using its weights computed from the validation set, the weighted mean is calculated for each nowcast in the testing weeks. The 95\% prediction intervals are computed similarly as above.

\section*{Acknowledgements}

We acknowledge Descartes Labs for remote sensing data and the Ministry of Health of Brazil for reported case counts data. Research presented in this article was supported by the Laboratory Directed Research and Development program of Los Alamos National Laboratory 20190581ECR, and 20180740ER. The content is solely the responsibility of the authors and does not necessarily represent the official views of the sponsors. SDV was supported in part by NIH/NIGMS grant R01-GM130668-02. The funders had no role in study design, data collection and analysis, decision to publish, or preparation of the manuscript. Los Alamos National Laboratory is operated by Triad National Security, LLC, for the National Nuclear Security Administration of U.S. Department of Energy (Contract No. 89233218CNA000001).

\section*{Author contributions statement}

KK wrote the manuscript, designed and implemented the statistical part of the study, analyzed the results, and prepared figures, KM cleaned and fused data and assisted with figures, AS did kriging for the data, JC helped plan the manuscript, fused data, and analyzed results, GF and AZ acquired and processed the remote sensing data and weather data and co-conceived the project, NP pulled GHT data, did data fusion, and assisted with computational methods, DO helped design the statistical methodology, NG provided case surveillance data and co-conceived the project, SDV co-conceived the project and led the team, CM co-conceived and coordinated the study, team, and methods. 
All authors reviewed the final manuscript.

\bibliography{sample}

\section*{Supplementary Material}

\subsection*{Figures}

\begin{figure}[ht]%{Satellite Variables by State 2}
\includegraphics[width=1\textwidth]{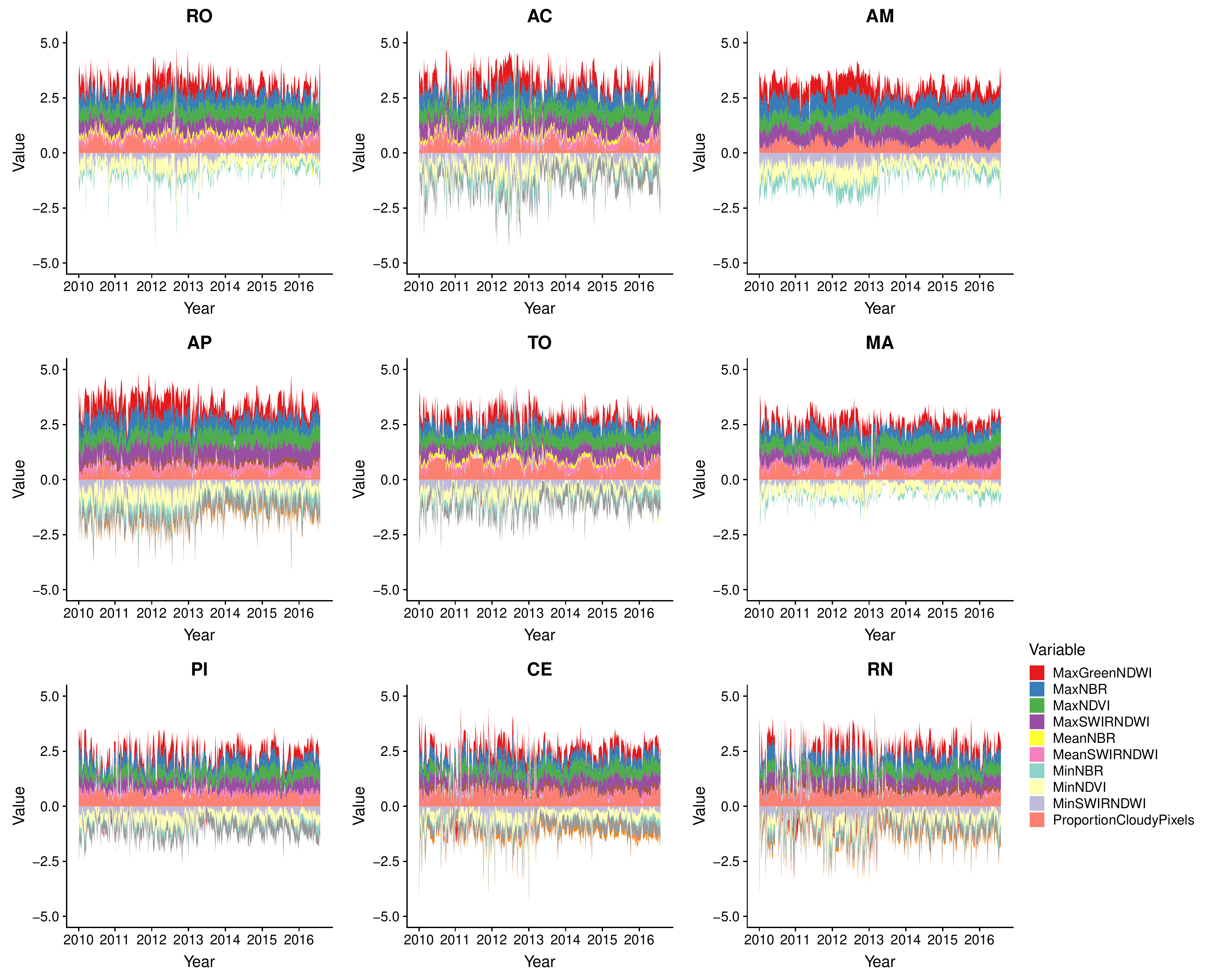}
\centering
\caption{\textbf{Satellite variables in each Brazilian state (2).}}
%\caption{For a selection of Brazilian states, satellite variables are plotted from 2010 to 2016. (Variables with approximately zero variance in all states are omitted.)} 
\label{fig:satellite2}
\end{figure}

\begin{figure}[ht]%{Satellite Variables by State 3}
\includegraphics[width=1\textwidth]{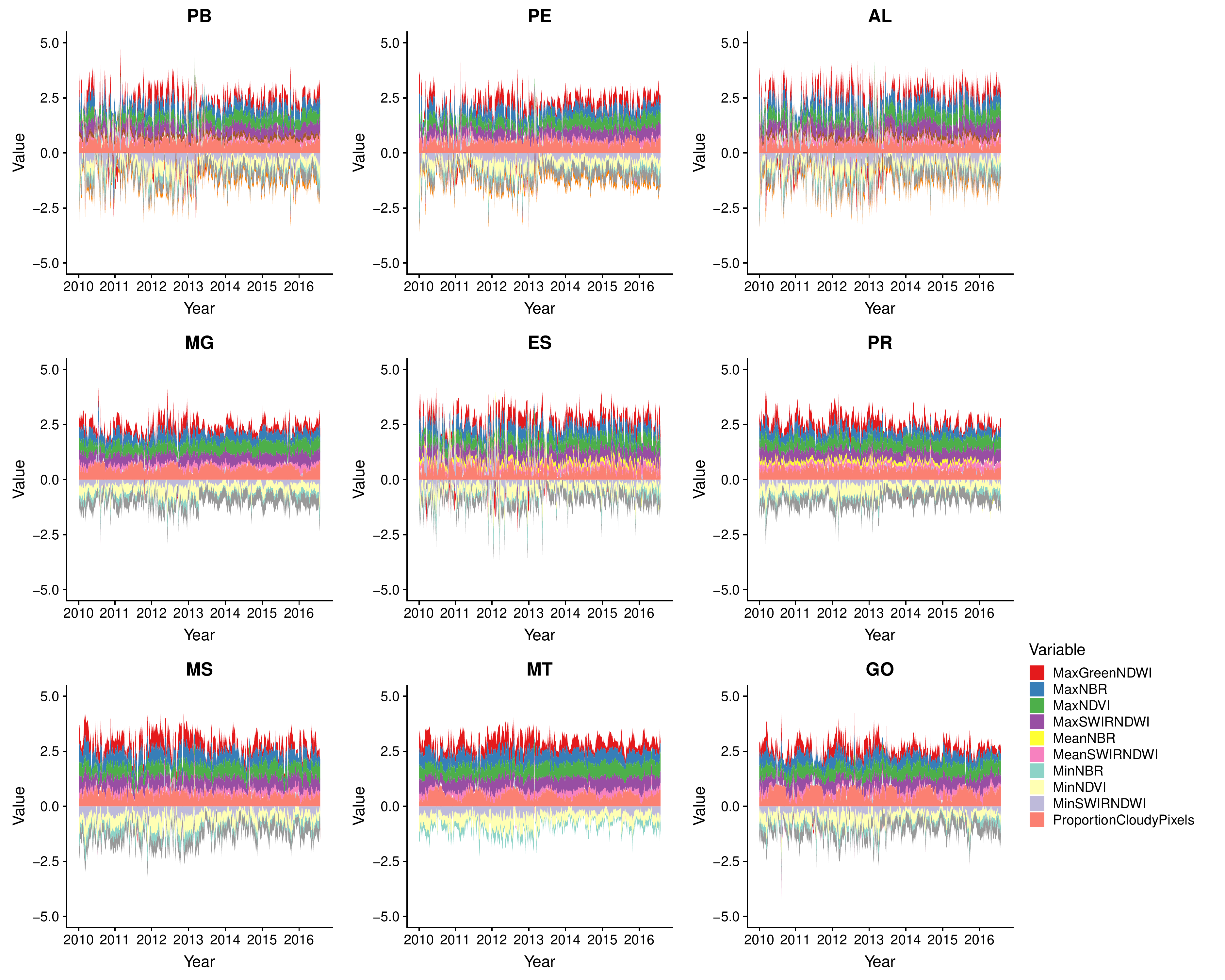}
\centering
\caption{\textbf{Satellite variables in each Brazilian state (3).}}
%\caption{For a selection of Brazilian states, satellite variables are plotted from 2010 to 2016. (Variables with approximately zero variance in all states are omitted.)} 
\label{fig:satellite3}
\end{figure}

\begin{figure}[ht]%{Weather Variables by State 2}
\includegraphics[width=1\textwidth]{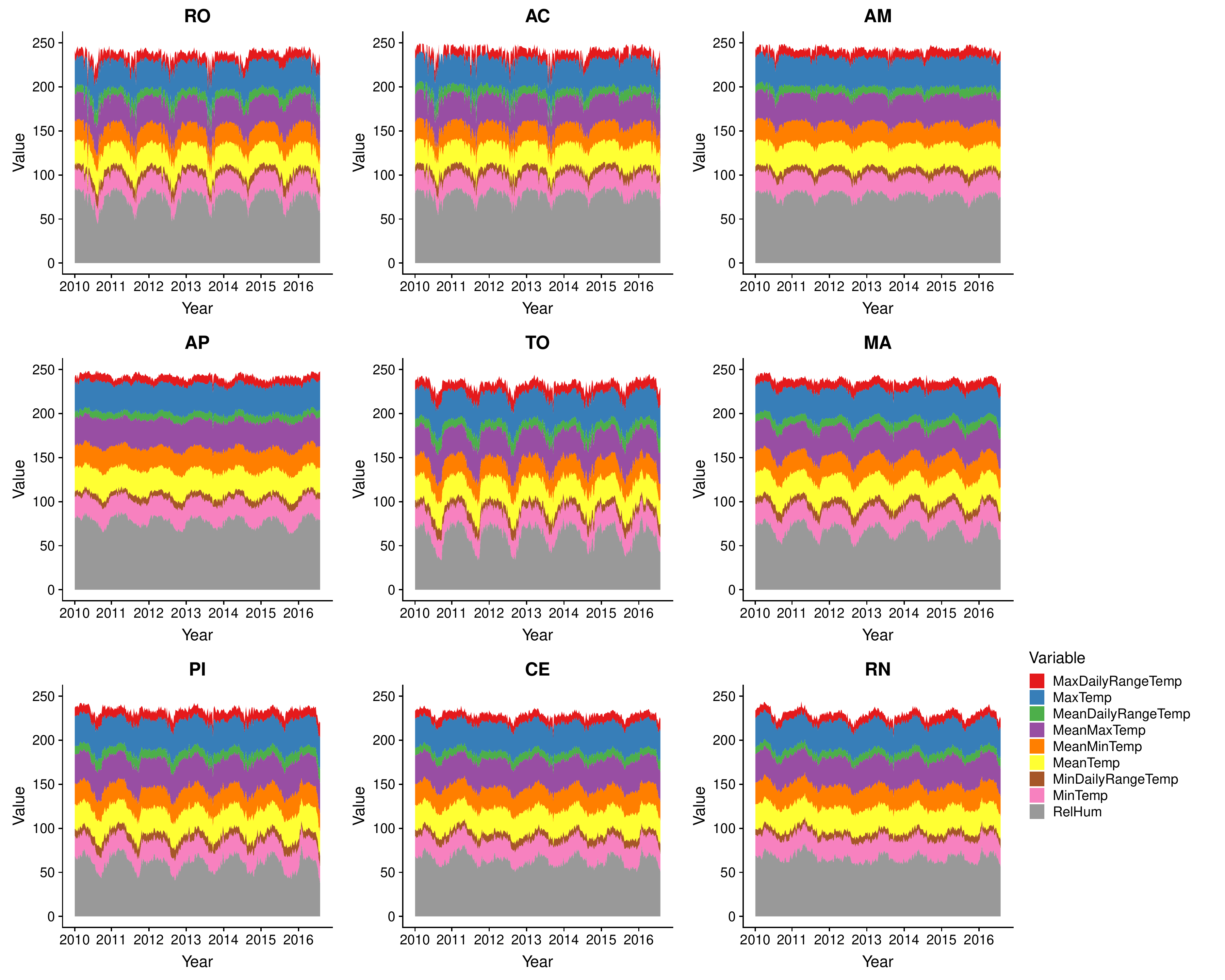}
\centering
\caption{\textbf{Weather variables in each Brazilian state (2).}}
%\caption{For a selection of Brazilian states, weather variables are plotted from 2010 to 2016. (Variables with approximately zero variance in all states are omitted.)} 
\label{fig:weather2}
\end{figure}

\begin{figure}[ht]%{Weather Variables by State 3}
\includegraphics[width=1\textwidth]{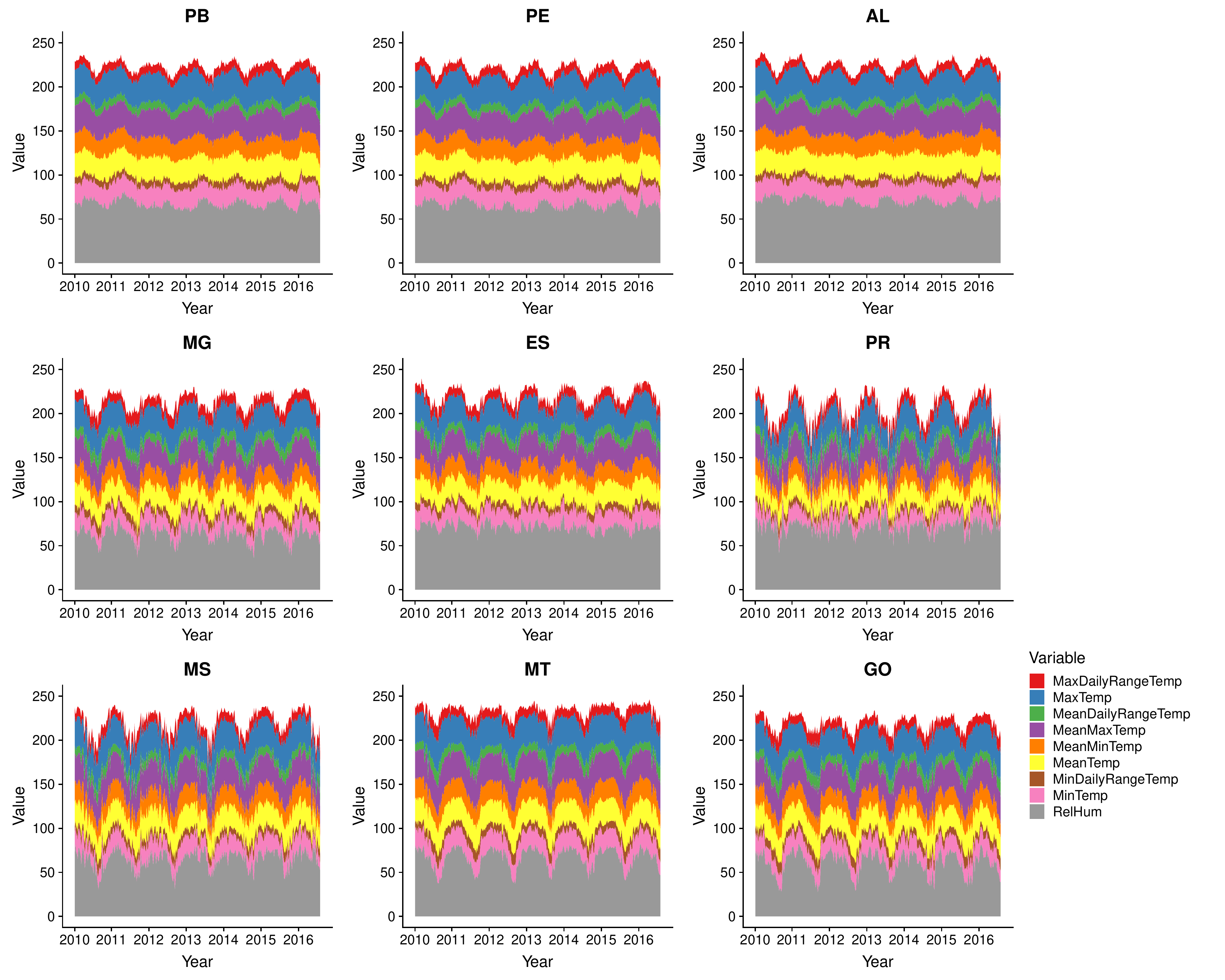}
\centering
\caption{\textbf{Weather variables in each Brazilian state (3).}}
%\caption{For a selection of Brazilian states, weather variables are plotted from 2010 to 2016. (Variables with approximately zero variance in all states are omitted.)} 
\label{fig:weather3}
\end{figure}

\begin{figure}[ht]%{Google Health Trends Variables by State 2 }
\includegraphics[width=1\textwidth]{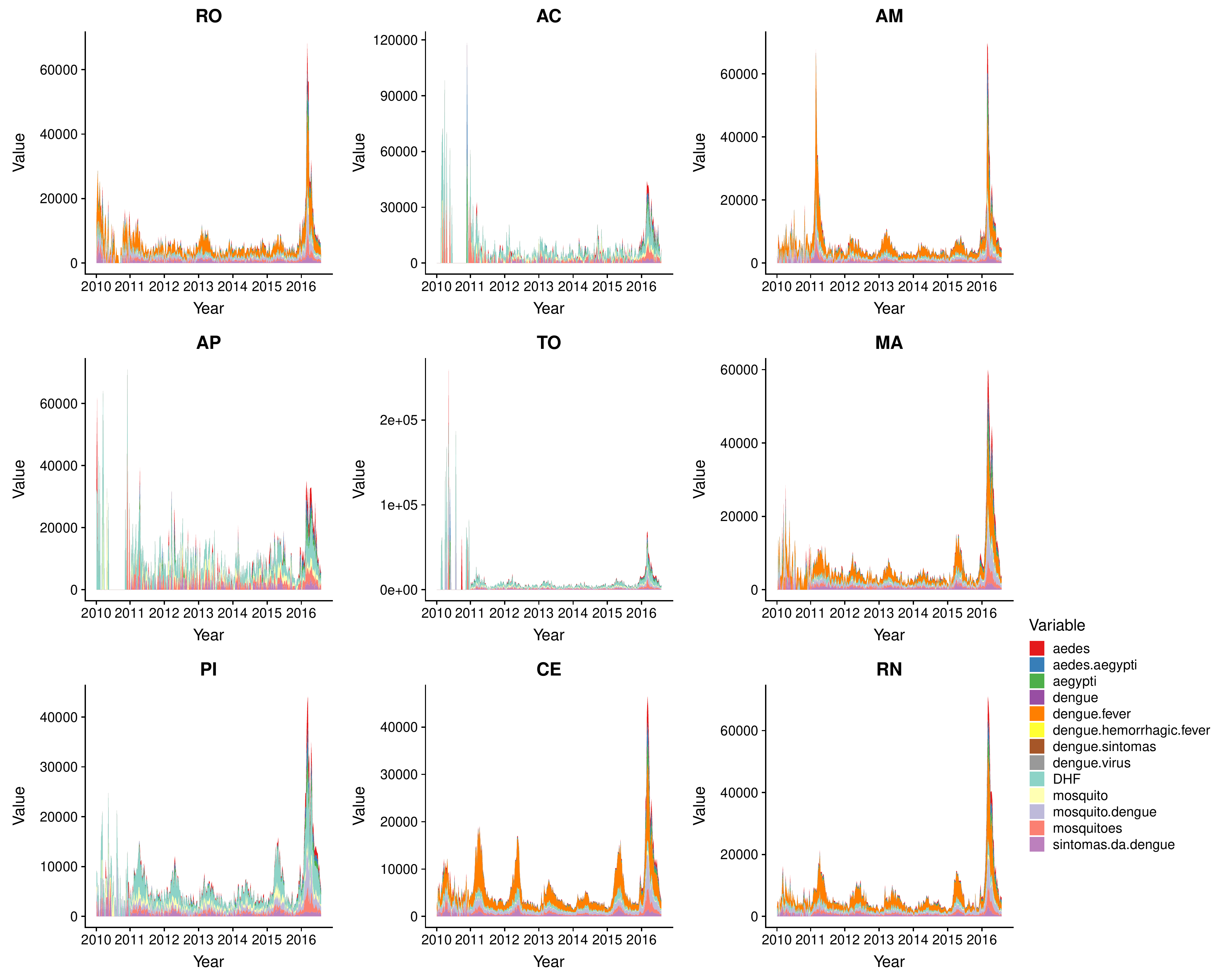}
\centering
\caption{\textbf{Google variables in each Brazilian state (2).}}
%\caption{For a selection of Brazilian states, Google Health Trends variables are plotted from 2010 to 2016. (Variables with approximately zero variance in all states are omitted.)} 
\label{fig:GHT2}
\end{figure}

\begin{figure}[ht]%{Google Health Trends Variables by State 3}
\includegraphics[width=1\textwidth]{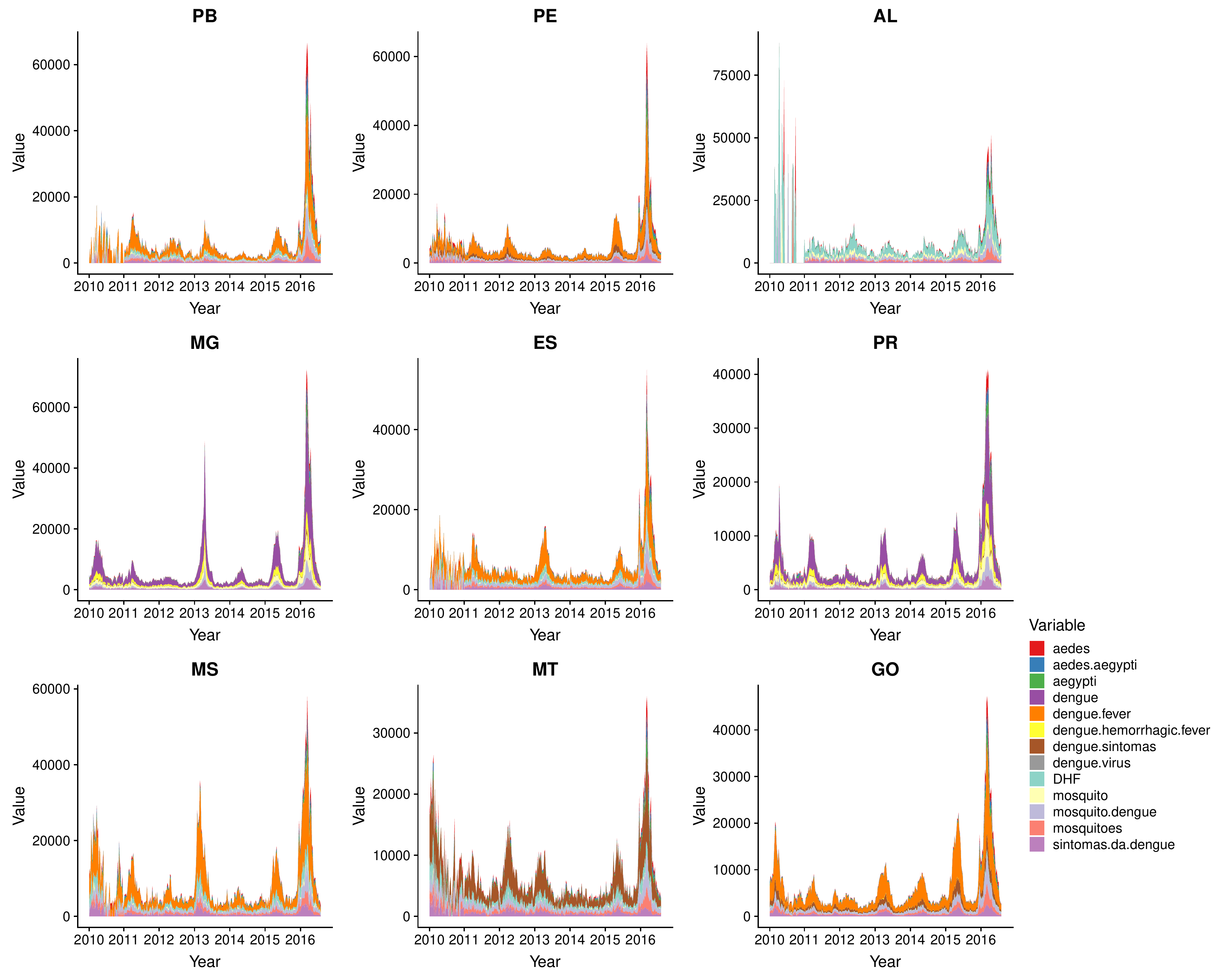}
\centering
\caption{\textbf{Google variables in each Brazilian state (3).}}
%\caption{For a selection of Brazilian states, Google Health Trends variables are plotted from 2010 to 2016. (Variables with approximately zero variance in all states are omitted.)} 
\label{fig:GHT3}
\end{figure}

\begin{figure}[ht]%{Dengue Clinical Cases Per Person 2}
\includegraphics[width=1\textwidth]{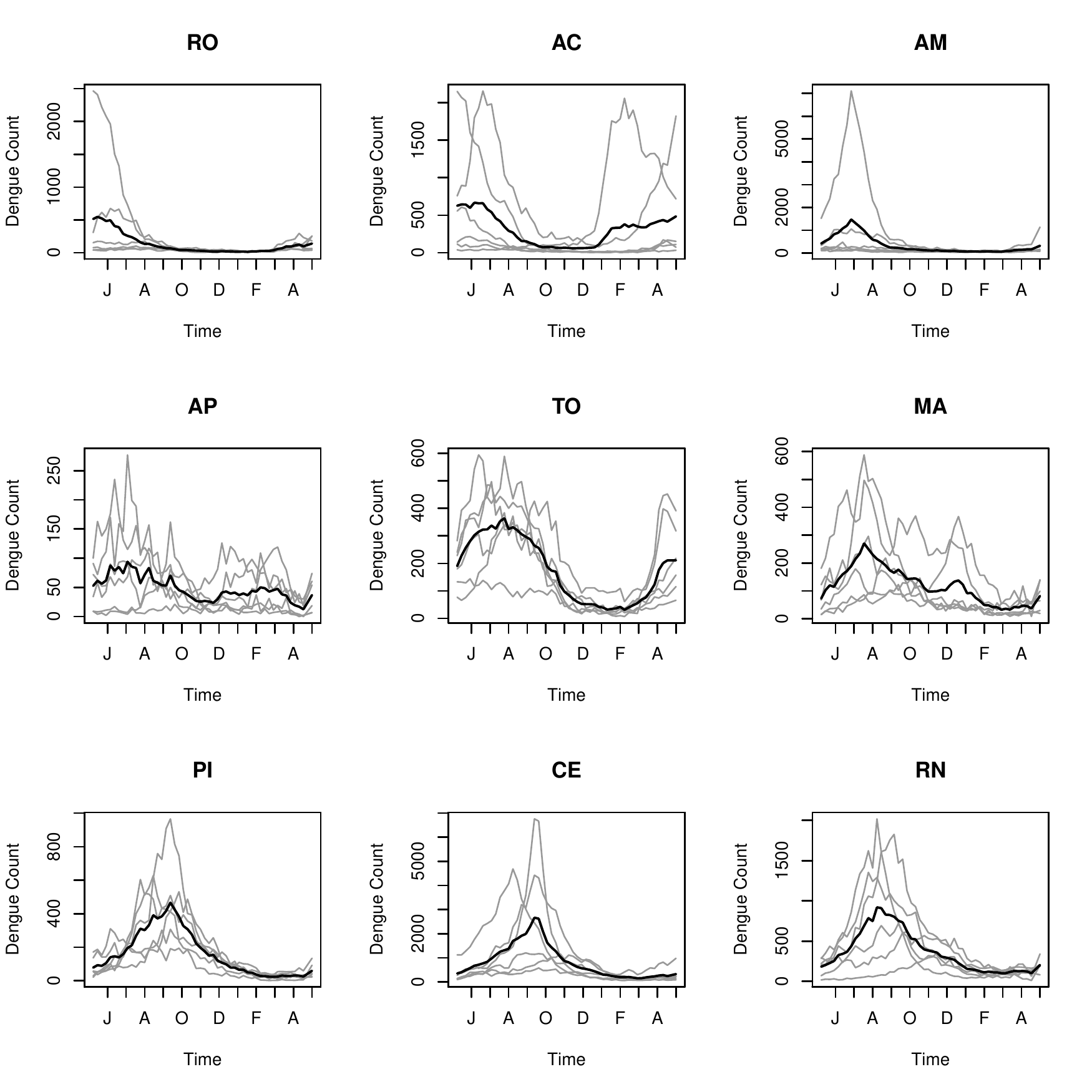}
\centering
\caption{\textbf{Mean dengue cases over epidemic weeks in each Brazilian state (2).}}
%\caption{For a selection of Brazilian states, the mean number of dengue cases per epidemic week (computed over all weeks from 2010 to 2015-6) is displayed.} 
\label{fig:cases_avg2}
\end{figure}

\begin{figure}[ht]%{Dengue Clinical Cases Per Person 3}
\includegraphics[width=1\textwidth]{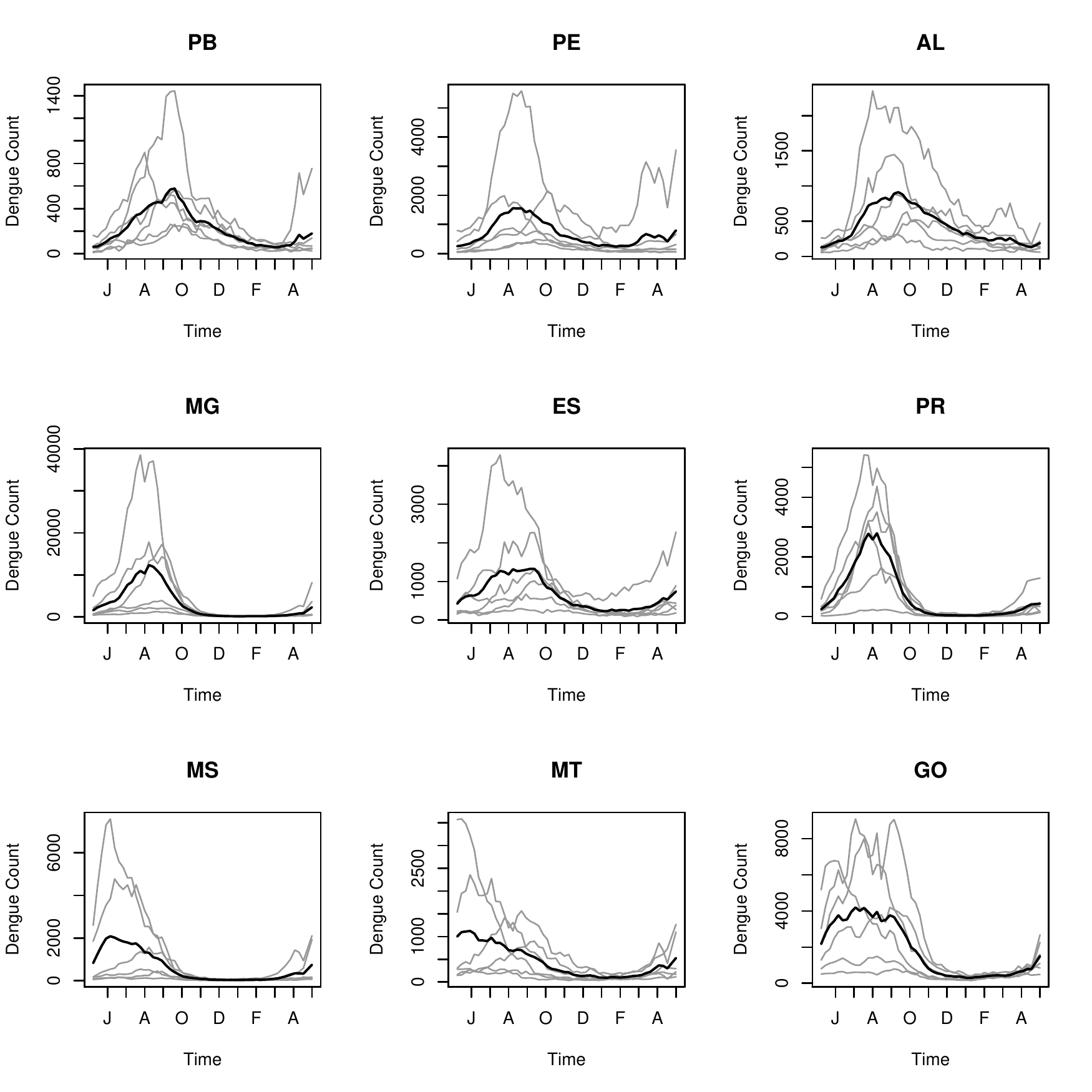}
\centering
\caption{\textbf{Mean dengue cases over epidemic weeks in each Brazilian state (3).}}
%\caption{For a selection of Brazilian states, the mean number of dengue cases per epidemic week (computed over all weeks from 2010 to 2015-6) is displayed.} 
\label{fig:cases_avg3}
\end{figure}

\begin{figure}[ht]%{Predicted vs. Observed Dengue by State 2}
\includegraphics[width=1\textwidth,trim={0cm 0cm 0cm 0cm},clip]{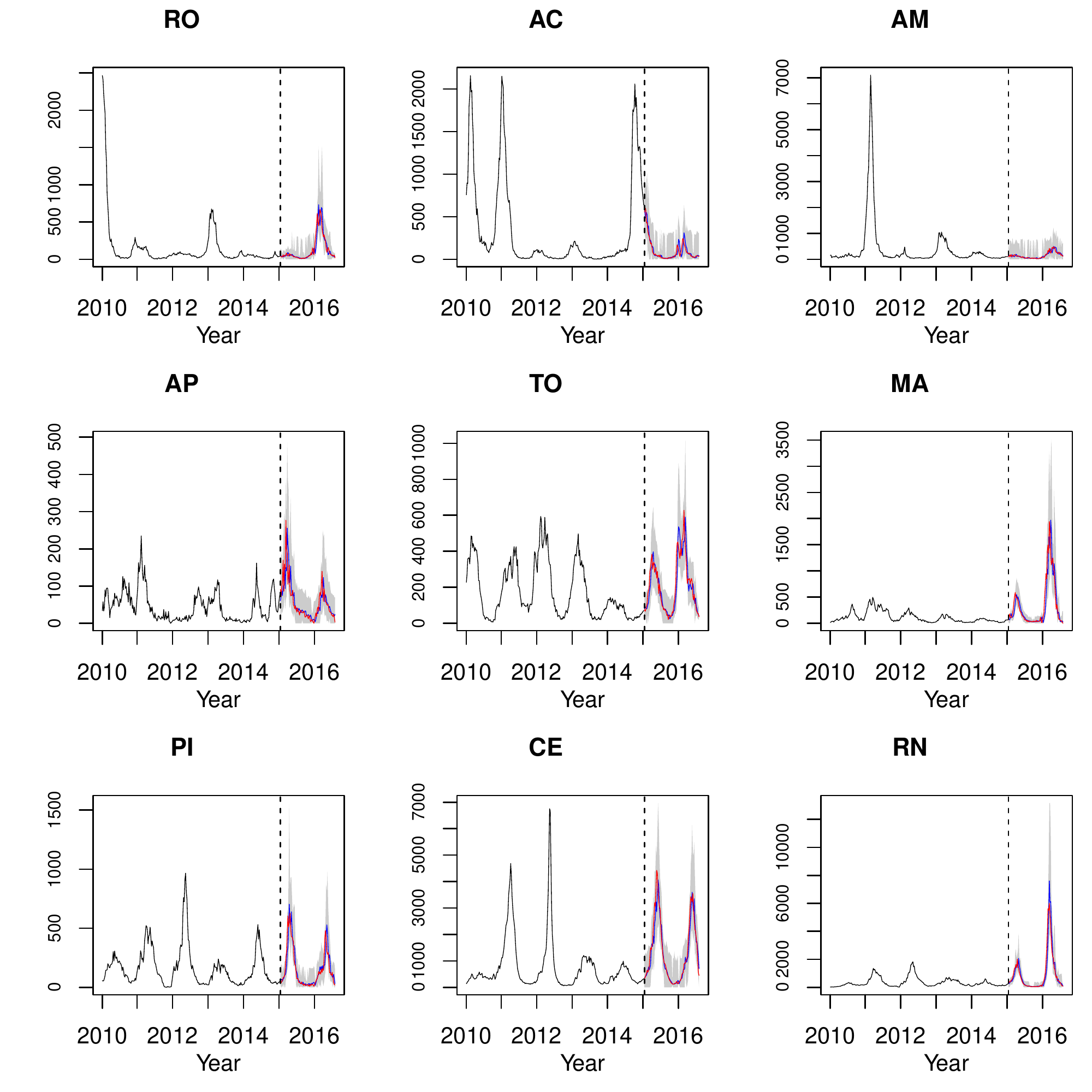}
\centering
\caption{\textbf{Observed vs. nowcasted dengue by Brazilian state from trimmed mean ensemble approach (2).}}
%\caption{Time series of recursively nowcasted vs. observed dengue are displayed for a selection of Brazilian states. The model varies by state and is chosen from model selection within the training weeks. The vertical dashed line marks the boundary between training and testing weeks. The black time series corresponds to observed dengue in the training weeks (years 2010-4). The blue and red lines correspond to predicted and observed dengue, respectively, within the testing weeks (years 2015-16). The gray shadings in the testing weeks represent the 95\% prediction intervals.} 
\label{fig:forecasts2}
\end{figure}

\begin{figure}[ht]%{Predicted vs. Observed Dengue by State 3}
\includegraphics[width=1\textwidth,trim={0cm 0cm 0cm 0cm},clip]{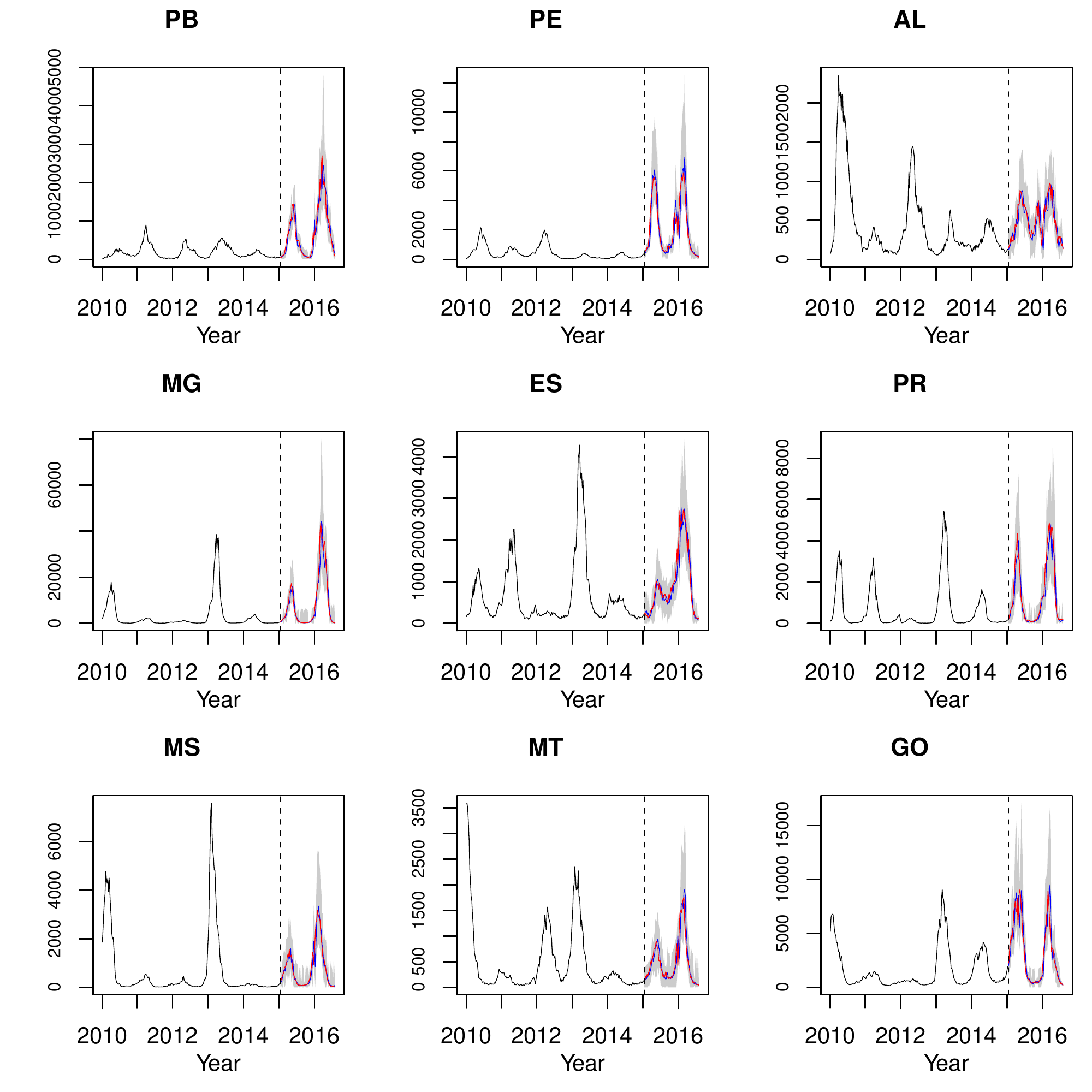}
\centering
\caption{\textbf{Observed vs. nowcasted dengue by Brazilian state from trimmed mean ensemble approach (3).}}
%\caption{Time series of recursively nowcasted vs. observed dengue are displayed for a selection of Brazilian states. The model varies by state and is chosen from model selection within the training weeks. The vertical dashed line marks the boundary between training and testing weeks. The black time series corresponds to observed dengue in the training weeks (years 2010-4). The blue and red lines correspond to predicted and observed dengue, respectively, within the testing weeks (years 2015-16). The gray shadings in the testing weeks represent the 95\% prediction intervals.} 
\label{fig:forecasts3}
\end{figure}

\end{document}